\documentclass[%
reprint,
amsmath,
amssymb,
aps,
prl,
nolongbibliography
]{revtex4-2}

\usepackage{graphicx}
\usepackage{dcolumn}
\usepackage{bm}

\usepackage{mathtools}

\usepackage{color}

\newcommand{\nc}{\newcommand}
\nc{\bra}{\langle}
\nc{\ket}{\rangle}
\nc{\vac}{|0\ket}
\nc{\da}{^{\dagger}}
\nc{\HASEP}{\mathcal{H}}
\nc{\im}{\text{i}}
\nc{\Imath}{\mathcal{I}}
\nc{\blue}{\textcolor{blue}}

\nc{\la}{\lambda}
\nc{\al}{\alpha}
\nc{\ze}{\zeta}
\nc{\laA}{\lambda_A}
\nc{\laB}{\lambda_B}
\nc{\laAB}{\lambda_{AB}^{nm}}

\begin{document}

\title{Asymmetry-induced delocalization transition in the integrable non-Hermitian spin chain}

\author{Yuki Ishiguro$^1$}
\author{Jun Sato$^2$}
\author{Katsuhiro Nishinari$^3$}
 \affiliation{$^1$Department of Aeronautics and Astronautics, Faculty of Engineering, The University of Tokyo, 7-3-1 Hongo, Bunkyo-ku, Tokyo 113-8656, Japan\\
 $^2$Faculty of Engineering, Tokyo Polytechnic University, 5-45-1 Iiyama-minami, Atsugi, Kanagawa 243-0297, Japan\\
 $^3$Research Center for Advanced Science and Technology, The University of Tokyo, 4-6-1 Komaba, Meguro-ku, Tokyo 153-8904, Japan
 }

\begin{abstract}
The emergence of quasiparticles is a universal property in integrable systems.
String-type quasiparticles, which are characterized by the string solutions of Bethe equations, play fundamental roles in the analysis of their physics.
Through an investigation of the Bethe equations in the asymmetric simple exclusion process, we reveal the existence of string solutions in the presence of non-Hermiticity resulting from asymmetrical hopping. Because of the non-Hermiticity, the string solutions exhibit exotic properties such as the complexification of the center of string solutions and the delocalization of Bethe quantum numbers. In addition, we find the picture of string-type quasiparticles collapses in the strong asymmetry regime. The collapse of string solutions characterizes the transition of eigenstates from bound states to scattering states.
\end{abstract}
\maketitle


\textit{Introduction.}---
Non-Hermitian physics has attracted attention as a promising approach for investigating general principles in non-equilibrium systems \cite{el2018non,ashida2020non}. 
Since non-Hermiticity effectively describes non-equilibrium phenomena, such as asymmetric hopping and dissipation-inflow of particles, non-Hermitian Hamiltonians appear in various situations describing systems out of equilibrium including self-driven particle systems \cite{derrida1998exactly,golinelli2006asymmetric,blythe2007nonequilibrium} and open quantum systems \cite{PhysRevA.79.023614,PhysRevLett.126.110404,PhysRevLett.68.580}.
Quantum integrable systems play crucial roles in understanding interacting many-body systems \cite{korepin1997quantum}.
Although examples of integrable non-Hermitian systems have been found \cite{derrida1998exactly,golinelli2006asymmetric,blythe2007nonequilibrium,PhysRevA.79.023614,PhysRevLett.126.110404}, the theoretical understanding for their analysis is less advanced than that for Hermitian systems.

Integrable systems are usually analyzed in terms of quasiparticles.
For example, the Heisenberg spin chain is understood through quasiparticles characterized by the string solutions, which are specific solutions for the Bethe equations widely observed among integrable systems \cite{takahashi_1999}.
String-type quasiparticles, which are closely related to solitons in classical integrable systems \cite{PhysRevB.92.214427,wadachi1984classical,wadati1985quantum,PhysRevA.40.854}, allows understanding the physics of integrable systems such as anomalous transport \cite{PhysRevLett.125.070601,Bulchandani_2021}.
Moreover, through the so-called string hypothesis, they enable us to use the powerful analytical frameworks of integrable systems, including the thermodynamic Bethe ansatz (TBA) \cite{takahashi_1999,yang1969thermodynamics,takahashi1971one,PhysRevLett.82.1764,mossel2012generalized} and the generalized hydrodynamics (GHD) \cite{PhysRevX.6.041065,PhysRevLett.117.207201,10.21468/SciPostPhys.6.4.049,10.21468/SciPostPhysLectNotes.18,RevModPhys.93.025003}.
Therefore, understanding the quasiparticle picture in integrable non-Hermitian systems is the key to establishing cornerstones for analyzing them.

In this Letter, we elucidate the string-type quasiparticle picture in the presence of non-Hermiticity caused by asymmetric hopping and present the mechanism of delocalization transition based on this picture. 
We consider the asymmetric simple exclusion process (ASEP), which is an integrable non-Hermitian spin chain describing asymmetric random walks of self-driven particles with hardcore interactions \cite{derrida1998exactly,essler1996representations,blythe2007nonequilibrium,golinelli2006asymmetric,gwa1992bethe,kim1995bethe,Golinelli_2004,Golinelli_2005,prolhac2013spectrum,prolhac2014spectrum,prolhac2016extrapolation,prolhac2017perturbative,de2005bethe,deGier_2006,deGier_2008,PhysRevLett.107.010602,Wen_2015,Crampe_2015}.
The ASEP provides an excellent field for investigating non-equilibrium physics, such as the KPZ universality class \cite{bertini1997stochastic,takeuchi2018appetizer} and the boundary-induced phase transition \cite{blythe2007nonequilibrium}, and has a wide range of applications including traffic flow \cite{schadschneider2000statistical,schadschneider2010stochastic} and biophysics \cite{macdonald1968kinetics,klumpp2003traffic}. 
Although the ASEP is exactly analyzable by the Bethe ansatz \cite{golinelli2006asymmetric,gwa1992bethe,kim1995bethe,Golinelli_2004,Golinelli_2005,golinelli2006asymmetric,prolhac2013spectrum,prolhac2014spectrum,prolhac2016extrapolation,prolhac2017perturbative,de2005bethe,deGier_2006,deGier_2008,PhysRevLett.107.010602,Wen_2015,Crampe_2015}, the understanding of the properties of eigenstates is still developing because of the complexity of the Bethe equations \cite{golinelli2006asymmetric,gwa1992bethe,kim1995bethe,Golinelli_2004,Golinelli_2005,golinelli2006asymmetric,prolhac2013spectrum,prolhac2014spectrum,prolhac2016extrapolation,prolhac2017perturbative}.
Our main result is to show that the Bethe roots can be understood in terms of the string solutions.
Because of the non-Hermiticity induced by asymmetric hopping, the string solutions exhibit intriguing properties such as the complexification of the center of string solutions and the delocalization of Bethe quantum numbers. 
One of the most notable phenomena is that the picture of the string solutions collapses in the strong asymmetry regime. 
We reveal that the collapse of strings characterizes the transition of eigenstates from bound states to scattering states.

\textit{Model.}---
The ASEP is a continuous-time Markov process in a one-dimensional lattice defined by the following rule \cite{derrida1998exactly,golinelli2006asymmetric,blythe2007nonequilibrium}. Each particle moves to the nearest right (left) site with the hopping rate $p$ ($q$). 
Because of the hardcore interactions, each site contains only a single particle at most. Without loss of generality, we can set $p+q=1$ and $p\ge q$. 
When $q=0$, particles move in only one direction. In this case, the model is called the totally asymmetric simple exclusion process (TASEP).

The time evolution of the ASEP is described by the imaginary-time Schr\"{o}dinger equation
\begin{align}
    \frac{d}{dt} |P(t)\ket = \HASEP |P(t)\ket,
    \label{eq:master}
\end{align}
where $|P(t)\ket$ is the stochastic state vector of a system at time $t$. 
The Hamiltonian $\HASEP$ with $L$ sites is given by
\begin{align}
    \HASEP 
    = \sum_{j=1}^{L} \left[ pS_{j}^{+}S_{j+1}^{-}+qS_{j}^{-}S_{j+1}^{+} +S_j^z S_{j+1}^z-\frac{1}{4}\right],
    \label{eq:HASEP}
\end{align}
where $S^{x,y,z}$ are half of the Pauli matrices and $S^{\pm}:=S^x \pm \im S^y$ are the ladder operators.
Here, we consider periodic boundary conditions.
When hopping rates are symmetric ($p=q$), the Hamiltonian (\ref{eq:HASEP}) is equivalent to that of the Heisenberg spin-$1/2$  chain. 
The asymmetry of hopping rates, which drives current, makes the Hamiltonian non-Hermitian.

The Hamiltonian (\ref{eq:HASEP}) is exactly diagonalized through the Bethe ansatz \cite{golinelli2006asymmetric}.
Thus, we obtain the corresponding eigenstates if we identify the solutions of the Bethe equations
\begin{align}
\begin{split}
    z_j^L = (-1)^{N-1} \prod_{\ell=1}^N \frac{p-z_j +qz_\ell z_j}{p-z_\ell +qz_\ell z_j} \\ 
    \text{for} \quad j=1,2,\cdots,N,    
\end{split}
    \label{eq:BAEofASEP}
\end{align}
where $N$ is the number of particles. 
The energy eigenvalue $E$ is described in terms of the Bethe roots $\{z_j\}$ as
\begin{align}
     E = q\sum_{j=1}^N z_j +p\sum_{j=1}^N \frac{1}{z_j}-N.
    \label{eq:eigenHASEP}
\end{align}
Because of the difficulty of solving the Bethe equations (\ref{eq:BAEofASEP}), many efforts have been made to clarify the distribution of the Bethe roots \cite{golinelli2006asymmetric,gwa1992bethe,kim1995bethe,Golinelli_2004,Golinelli_2005,golinelli2006asymmetric,prolhac2013spectrum,prolhac2014spectrum,prolhac2016extrapolation,prolhac2017perturbative}.

\textit{String solutions.}---
Here, we reveal that the Bethe equations (\ref{eq:BAEofASEP}) have the string solutions.
First, we introduce new variables $\{\la_j\}$ as
\begin{align}
    z_j=\frac{1}{\al}\frac{\sin\zeta(\la_j+\frac{\im}{2})}{\sin\ze (\la_j-\frac{\im}{2})},
    \label{eq:rapidity}
\end{align}
where $\alpha := \sqrt{q/p}$ and $\ze := -\log\al$. Then, the Bethe equations (\ref{eq:BAEofASEP}) are written as
\begin{align}
\begin{split}
            \left[ \frac{1}{\al}\frac{\sin\ze(\la_j+\frac{\im}{2})}{\sin\ze (\la_j-\frac{\im}{2})} \right]^L =  \prod_{\ell \neq j}^N \frac{\sin\ze(\la_j - \la_\ell + \im)}{\sin\ze(\la_j - \la_\ell - \im)} \\
            \text{for} \quad j=1,2,\cdots,N,    
\end{split}
        \label{eq:NewBAEASEP}
\end{align}
and the energy eigenvalue $E$ (\ref{eq:eigenHASEP}) is given by
\begin{align}
    E= - \frac{1}{2} \tanh \ze \sum_{j=1}^N a_1(\la_j), 
     \label{eq:energynewnotation}
\end{align}
where we introduce functions $a_n(\la)$ ($n\in \mathbb{N}$)
\begin{align}
    \begin{split}
        a_n(\la):=
        \begin{dcases}
        \frac{2 \sinh n \ze}{\cosh n\ze -\cos 2\ze \la}    & \text{for} \quad \alpha \neq 1 \\
        \frac{4n}{4 \la^2 +n^2}  & \text{for} \quad \alpha = 1.
        \end{dcases}
    \end{split}
    \label{eq:defa}
\end{align}
Note that we cannot define the transformation (\ref{eq:rapidity}) in the TASEP ($q=0$).
Therefore, it is difficult to formulate string solutions in the case of the TASEP.
In the following, we consider $q \neq 0$.

The new Bethe equations (\ref{eq:NewBAEASEP}) allow the formulation of the string solutions. We denote a Bethe root as $\la = a + \im b$ ($a,b \in \mathbb{R}$).
The $|\text{LHS}|^{1/L}$ of Eqs. (\ref{eq:NewBAEASEP}) is classified into three types according to the relation between the imaginary and real parts of the Bethe root as follows \cite{Supple}:
\begin{align}
   \left| \frac{1}{\al}\frac{\sin\ze(\la+\frac{\im}{2})}{\sin\ze (\la-\frac{\im}{2})} \right| \begin{dcases}
        >1 & \quad \text{for} \quad b > c_1(a) \\
        =1 & \quad \text{for} \quad b = c_1(a) \\
        <1 & \quad \text{for} \quad b < c_1(a),
    \end{dcases}
    \label{eq:lhscases}
\end{align}
where we introduce functions $c_n(x)$ ($n \in \mathbb{N}$) as
\begin{align}
    c_n(x)=\frac{1}{2\ze} \log \frac{\cos 2\ze x}{\cosh n \ze}.
    \label{eq:centerfunction}
\end{align}
By considering the $L \to \infty$ limit with fixed $N$, we obtain the string solutions. 
We assume the Bethe root $\la_j=a_j + \im b_j$ with $b_j > c_1(a_j)$. 
From Eqs. (\ref{eq:lhscases}), the $|\text{LHS}|$ of Eq. (\ref{eq:NewBAEASEP}) diverges in the $L\to \infty$ limit.
From the consistency of the Bethe equations (\ref{eq:NewBAEASEP}), the denominator of Eq. (\ref{eq:NewBAEASEP}) goes to $0$ in this limit. 
This implies that there exists an integer $\ell$ ($1 \le \ell \le N,$ $\ell \neq j$) such that $\la_\ell = \la_j - \im$ in $L\to  \infty$. Thus, if the Bethe root $\la_j$ with $b_j > c_1(a_j)$ exists, another Bethe root exists at the position displaced from $\la_j$ by $-\im$. Similarly, the Bethe root $\la_j$ with $b_j < c_1(a_j)$ generates another Bethe root at the position displaced from $\la_j$ by $\im$.
From this discussion, we obtain a series of Bethe roots 
\begin{align}
\begin{split}
     \la_A^{n,j} = \la_A^n + \frac{\im}{2}(n+1-2j) + \delta_A^{n,j}
     \\ \text{for} \quad j=1,2,\cdots,n.    
\end{split}
\label{eq:stringsolutions}
\end{align}
These roots are called $n$-string solutions. $A$ is the index of the string, $n$ is the length of the string, $\delta_A^{n,j}$ is the deviation that vanishes when $L \to \infty$, and $\la_A^n$ is called string center.
The difference from the standard string solutions is that the string center $\la_A^n$ is no longer necessarily real.
The string center in the Heisenberg spin chain is restricted to real because of the self-conjugacy of the Bethe roots \cite{vladimirov1986proof}. 
Since the non-Hermiticity violates the self-conjugacy, the string center is allowed to have an imaginary part, which is discussed in Supplemental Material (SM) \cite{Supple}.

The complexification of the string centers seems to make the analysis difficult due to the increased degrees of freedom.
However, the string centers of the ASEP have specific relations between the real and imaginary parts.
Therefore, the degree of freedom of the position of string centers remains one in the complex plane.
The relations are obtained by analyzing the Bethe-Takahashi equations, which are the reduced Bethe equations in terms of strings.
To derive the Bethe-Takahashi equations, we express all Bethe roots by the string solutions (\ref{eq:stringsolutions}). 
That is, $N$ Bethe roots are partitioned as $\sum_{n=1}^N n N_n =N$ where $N_n$ is the number of $n$-strings.
Then, the Bethe equations (\ref{eq:NewBAEASEP}) on an $n$-string are written as
\begin{align}
\begin{split}
            \left[ \frac{1}{\al}\frac{\sin\ze(\laA^{n,j}+\frac{\im}{2})}{\sin\ze (\laA^{n,j}-\frac{\im}{2})} \right]^L =  \prod_{\substack{(m,B) \\ \neq (n,A)}} \prod_{k= 1}^m \frac{\sin\ze(\laA^{n,j} - \laB^{m,k} + \im)}{\sin\ze(\laA^{n,j} - \laB^{m.k} - \im)} \\ 
            \times \prod_{j'\neq j}^n \frac{\sin\ze(\laA^{n,j} - \laA^{n,j'} + \im)}{\sin\ze(\laA^{n,j} - \laA^{n.j'} - \im)} \quad \text{for} \quad j=1,2,\cdots, n.
        \label{eq:BAEstringform}
\end{split}
\end{align}
By multiplying the Bethe equations of the $n$-string (\ref{eq:BAEstringform}) and assuming the deviations $\delta_A^{j,n}$ vanish, we obtain the Bethe-Takahashi equations \cite{Supple}
\begin{align}
    \left[ \frac{1}{\al^n} \frac{\sin\ze(\laA^{n}+\frac{\im}{2}n)}{\sin\ze (\laA^{n}-\frac{\im}{2}n)} \right]^L = \prod_{\substack{(m,B)\\ \neq (n,A)}} \varPhi_{nm}(\la_A^{n}-\la_B^{m}),
    \label{eq:BTeqs}
\end{align}
where $\varPhi_{nm}(\la):=\frac{\sin \ze (\la+\frac{\im}{2}|n-m|)}{\sin \ze (\la-\frac{\im}{2}|n-m|)} \left[ \frac{\sin \ze [\la+\frac{\im}{2}(|n-m|+2)]}{\sin \ze [\la-\frac{\im}{2}(|n-m|+2)]} \right]^2 \cdots$ $\left[ \frac{\sin \ze [\la+\frac{\im}{2}(n+m-2)]}{\sin \ze [\la-\frac{\im}{2}(n+m-2)]} \right]^2 \frac{\sin \ze [\la+\frac{\im}{2}(n+m)]}{\sin \ze [\la-\frac{\im}{2}(n+m)]}$ for $n \neq m$ and $\varPhi_{nm}(\la):=\left[ \frac{\sin \ze (\la+\im)}{\sin \ze (\la-\im)} \right]^2  \cdots  \left[ \frac{\sin \ze [\la+\im(n-1)]}{\sin \ze [\la-\im(n-1)]} \right]^2 \frac{\sin \ze (\la+\im n)}{\sin \ze (\la-\im n)}$ for $n=m$.
Equations (\ref{eq:BTeqs}) reveal the properties of the string centers. 
If we obtain the string centers, the energy eigenvalue (\ref{eq:energynewnotation}) is calculated from 
\begin{align}
E=- \frac{1}{2} \tanh \ze \sum_{n,A} a_n(\la_A^n).
\label{eq:energycenter}
\end{align}

In the following, we derive the relation between the real and imaginary parts of the string centers.
We denote a string center as $\la_A^n=a_A^n + \im b_A^n$  ($a_A^n,b_A^n \in \mathbb{R}$).
The $|\text{LHS}|^{1/L}$ of Eqs. (\ref{eq:BTeqs}) is classified into three types according to the relation between the imaginary and real parts of the string center \cite{Supple}:
\begin{align}
   \left| \frac{1}{\al^n}\frac{\sin\ze(\la+\frac{\im}{2}n)}{\sin\ze (\la-\frac{\im}{2}n)} \right| \begin{dcases}
        >1 & \quad \text{for} \quad b_A^n > c_n(a_A^n) \\
        =1 & \quad \text{for} \quad b_A^n = c_n(a_A^n) \\
        <1 & \quad \text{for} \quad b_A^n < c_n(a_A^n).
    \end{dcases}
    \label{eq:BTlhscases}
\end{align}
As we discussed in the derivation of the string solutions (\ref{eq:stringsolutions}), if the string center $\la_A^n$ with $b_A^n > c_n(a_A^n)$ ($b_A^n < c_n(a_A^n)$) exists, another string center $\la_B^m = \la_A^n - \frac{\im}{2} \ell$ ($\la_B^m = \la_A^n + \frac{\im}{2} \ell$) also exists in the $L \to \infty$ limit, where $\ell = |n-m|$, $|n-m|+2$,  $\cdots$, or $n+m$.
However, this contradicts the definition of the string center, which is a representative point of a line parallel to the imaginary axis.
This implies that the string centers have to satisfy the relation $b_A^n = c_n(a_A^n)$. Since $b_A^n$ is real, $a_A^n$ is restricted to $-\frac{\pi}{4 \zeta} < a_{A}^n < \frac{\pi}{4 \zeta}$.
Therefore, the string solutions are expressed by a real number $a_{A}^n$ ($-\frac{\pi}{4 \zeta} < a_{A}^n < \frac{\pi}{4 \zeta}$) as 
\begin{align}
    \begin{split}
        \la_A^{n,j} = a_{A}^n + \im \left[ \frac{n+1-2j}{2} + c_n(a_{A}^n) \right] + \delta_A^{n,j}
     \\ \text{for} \quad j=1,2,\cdots,n.    
    \end{split}
    \label{eq:realstringsolutions}
\end{align}

\begin{figure}[thb]
    \centering
    \includegraphics[height=12.5cm]{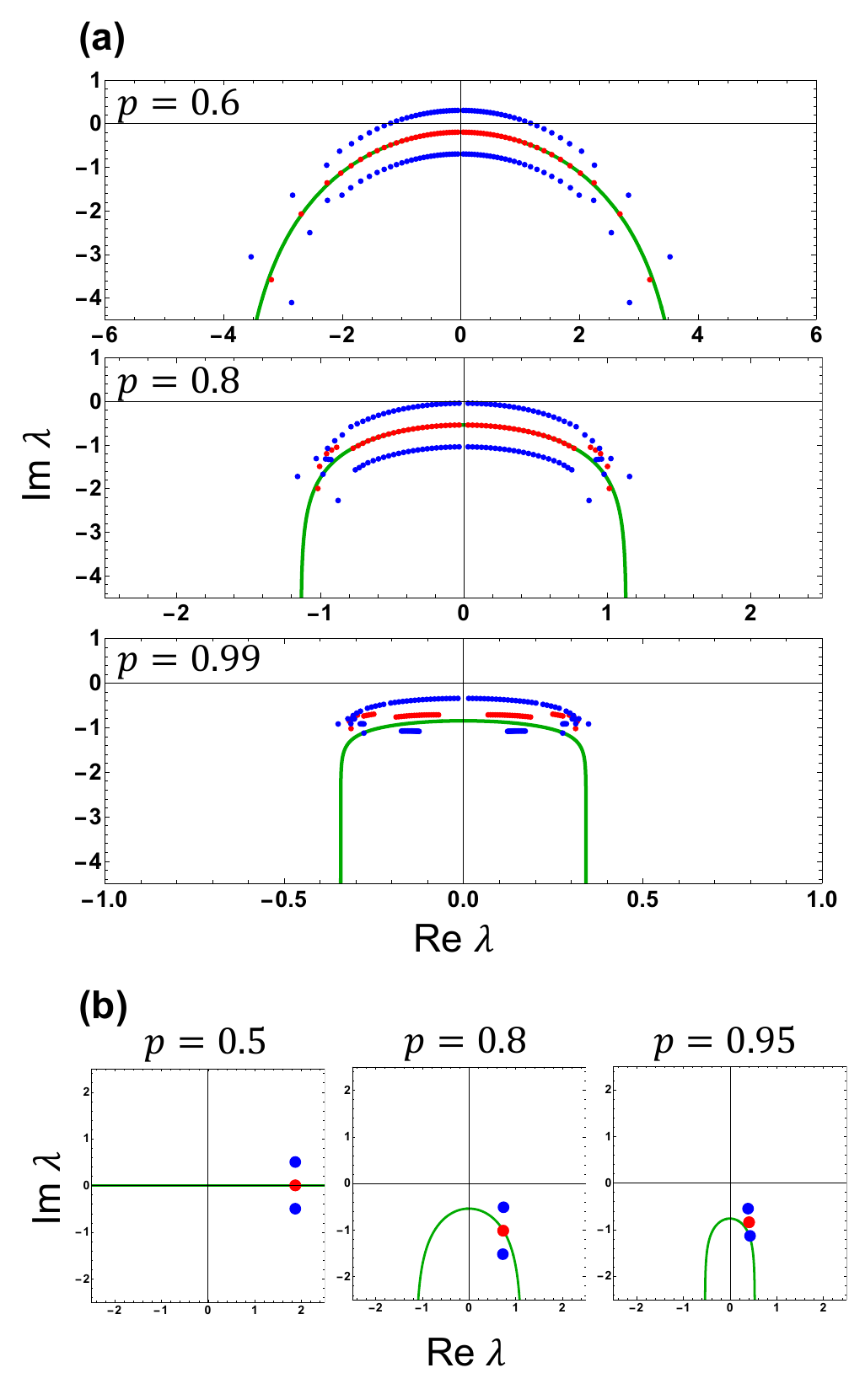}
    \caption{Two-string solutions of the ASEP for $L = 64$ and $N = 2$. (a) Distribution of the two-string solutions. (b) An example of the collapse of the two-string. Blue dots show Bethe roots, red dots show string centers, and green curves show the relation between the real and imaginary parts of string centers ($b_A^2 = c_2(a_A^2)$).}
    \label{fig:numstring}
\end{figure}

\textit{Collapse of strings.}---
To confirm the validity of the string solutions, we numerically solve the Bethe equations of the ASEP \cite{Supple}.
Although the string solutions are derived in the $L \to \infty$ limit, we can find them even in finite systems. We show the numerical results of the two-string solutions in Fig. \ref{fig:numstring}.
We confirm that the string solutions with complex string centers exist except near the TASEP ($p=1$). 
Figure \ref{fig:numstring} (a) shows that the Bethe roots are distributed approximately $\im$ apart on a line parallel to the imaginary axis, and the string center satisfies the relation $b_A^2 = c_2(a_A^2)$.
Conversely, in the strong asymmetry regime, the picture of the string solutions collapsed.
In other words, the difference between the Bethe roots is less than $\im$, and the string center does not satisfy $b_A^2 = c_2(a_A^2)$ near the TASEP (Fig. \ref{fig:numstring} (b)).
The cause of the collapse of strings is explained as follows.
As discussed before, the Bethe equations of the TASEP do not have string solutions. 
Since we consider finite systems in the numerical calculations, we observe the effect of the TASEP near $p=1$.
In the $L \to \infty$ limit, the collapse of strings occurs only in the TASEP.
In addition, we find the intriguing phenomenon about the Bethe quantum number from the numerical analysis. 
Whereas the Bethe quantum numbers of the standard string solutions are localized \cite{Hagemans_2007}, those of the ASEP are not. The delocalization of Bethe quantum numbers is shown in SM \cite{Supple}.

\textit{Delocalization transition.}---
Here, we investigate the properties of eigenstates based on the string-type quasiparticle picture through the analysis of the two-body problem. 
In the case of the Heisenberg spin chain ($p=1/2$), the Bethe equations have two types of solutions: real solutions (one-string solutions) and complex solutions ($n$-string solutions ($n>2$)) \cite{karbach}. 
The real solutions correspond to scattering states, and the complex solutions to bound states \cite{karbach}.

\begin{figure}[tbh]
    \centering
    \includegraphics[height=8.5cm]{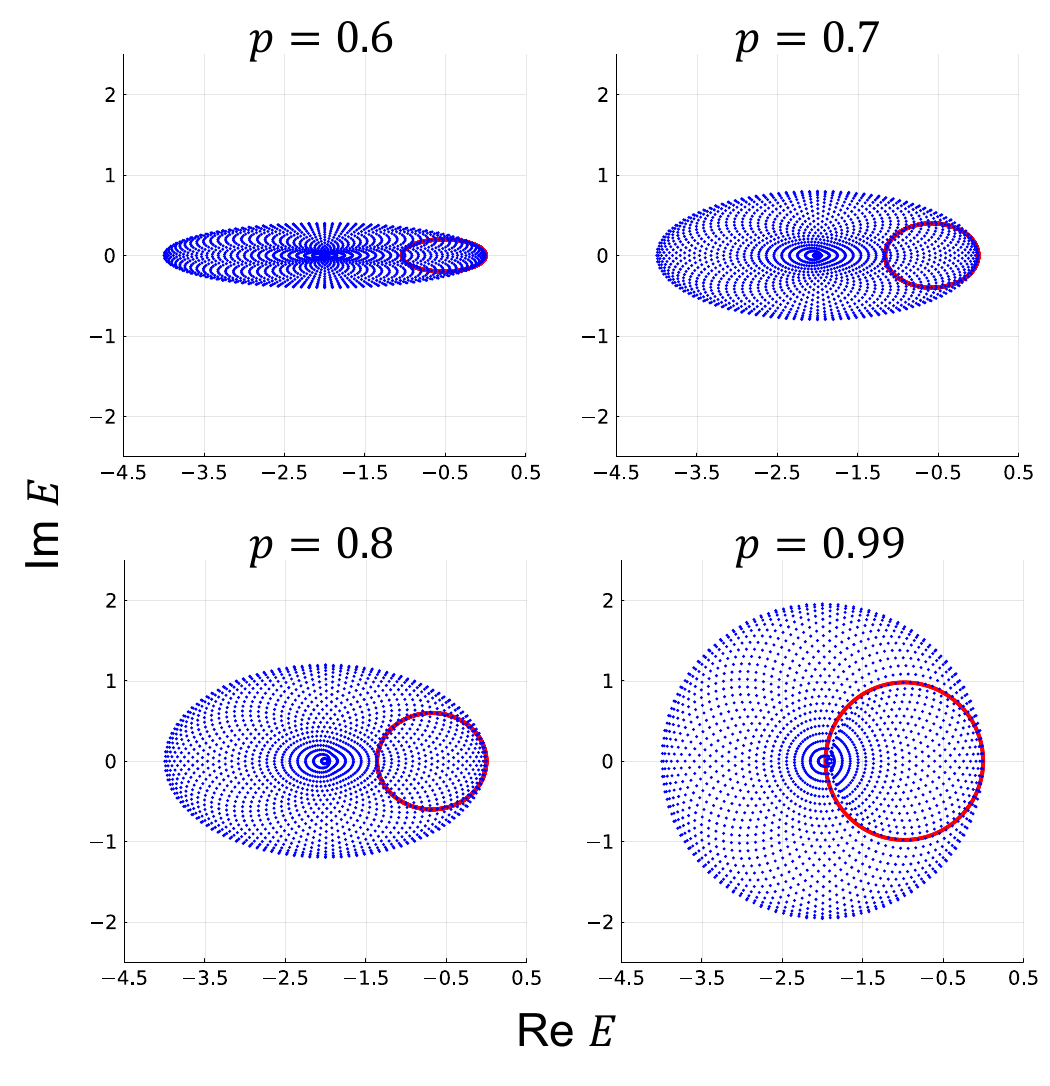}
    \caption{Complex spectra of the ASEP for $L=64$ and $N=2$. Blue dots show the spectra obtained by numerical diagonalization, and red curves show the spectra of two-string solutions (Eq. (\ref{eq:2stingspectra})).}
    \label{fig:spectra}
\end{figure}

Scattering states and bound states are easily distinguished by observing the dispersion relation in the Heisenberg spin chain \cite{Supple}.
Similarly, we find two types of complex spectra of the ASEP, except near $p=1$, as shown in Fig. \ref{fig:spectra}. One type of spectra is regularly distributed inside the large ellipse. The other is distributed on the small ellipse.
We call the former type-I and the latter type-II. 
The type-II spectra correspond to the two-string solutions.
From Eqs. (\ref{eq:energycenter}) and (\ref{eq:realstringsolutions}), the energy eigenvalues of the two-string solutions with the assumption that the deviations vanish ($\delta_{A}^{n,j}\to 0$) are analytically expressed by the real part of the string center $x$ ($-\frac{\pi}{4 \zeta} < x < \frac{\pi}{4 \zeta}$) as 
\begin{align}
E(x)=\frac{8\cosh 2\ze \sinh^2 \ze \cos^2 2\ze x}{\cos 4\ze x- \cosh 4 \ze}(1+\im \tan 2\ze x).
\label{eq:2stingspectra}
\end{align}
This describes an ellipse in a complex plane.
Figure \ref{fig:spectra} shows that the ellipse of Eq. (\ref{eq:2stingspectra}) coincides with the distribution of type-II spectra except in $p=0.99$.
Conversely, the spectra do not distribute on the ellipse of the two-string solutions (\ref{eq:2stingspectra}) near the TASEP where the strings collapse (Fig. \ref{fig:spectra} (p=0.99)). 
Generally, as the asymmetry increases, the ellipse of type-II spectra becomes larger. Near $p=1$, the ellipse of type-II spectra blends into the distribution of type-I spectra and becomes hard to distinguish.

\begin{figure}[tbh]
    \centering
    \includegraphics[height=8.5cm]{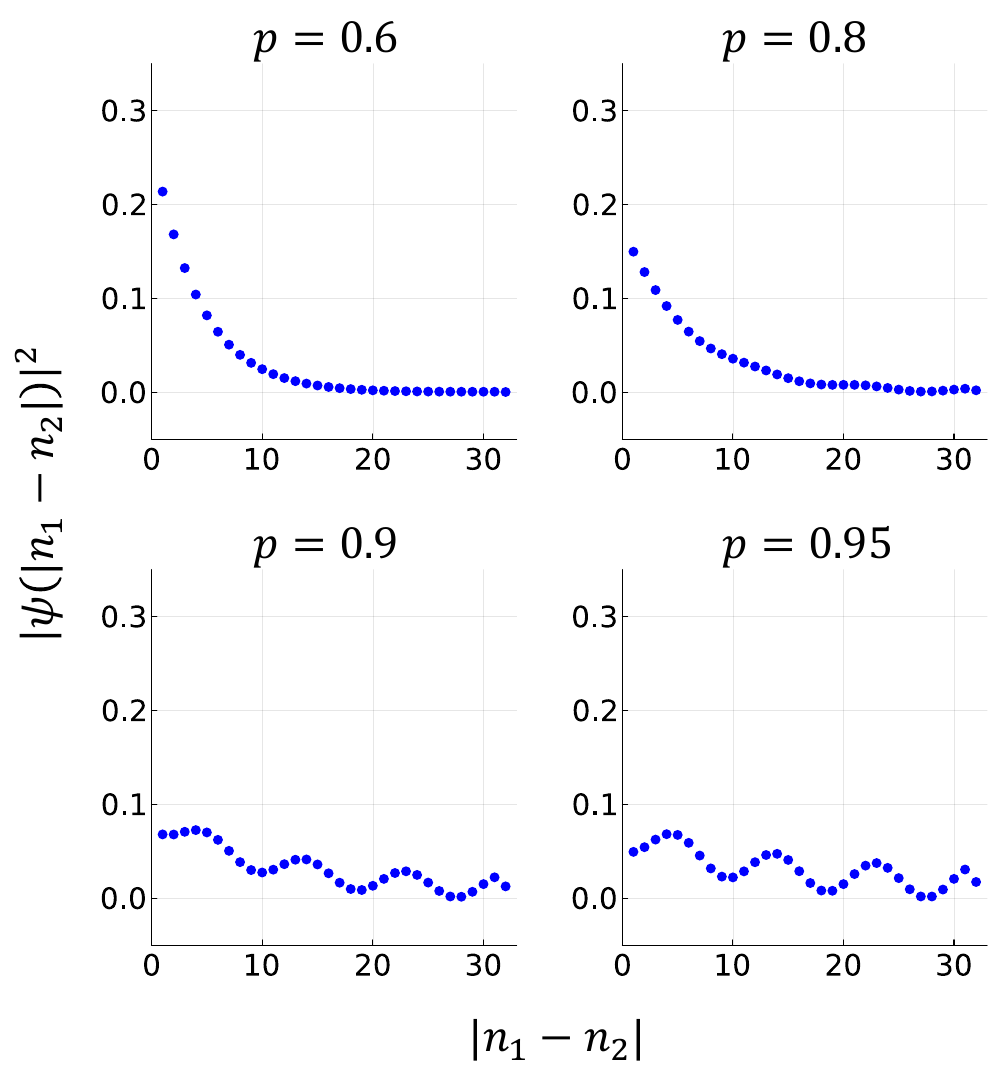}
    \caption{Weight distributions of the eigenstates of two-string solutions against the distance between particles in the ASEP for $L=64$ and $N=2$.}
    \label{fig:transition}
\end{figure}

In the Heisenberg spin chain, the complex solutions form bound states where particles tend to be localized. 
Similarly, the two-string solutions of the ASEP form bound states.
Figure \ref{fig:transition} shows the weight distributions of the eigenstates of two-string solutions against the distance between particles. 
The weight distribution is calculated as follows. 
We express an eigenstate $|\psi\ket$ by the basis vectors $|n_1,n_2\ket := S_{n_1}^- S_{n_2}^- |0\ket$ as $|\psi\ket = \sum_{1\le n_1 < n_2 \le N} \psi(n_1,n_2) |n_1,n_2\ket$, where $|0\ket$ is the vacuum.
The weight distribution is defined as $|\psi(\ell)|^2:=\sum_{|n_1-n_2|=\ell} |\psi(n_1,n_2)|^2$.
We find that the weight distribution increases as the distance between particles decreases except near the TASEP (Fig. \ref{fig:transition}).
Thus, the two-string solutions of the ASEP form bound states.
Conversely, in the strong asymmetry regime where the string solutions collapse, the localization of particles disappears.
This indicates that the eigenstate of the string solutions transition from the bound state to the scattering state through the collapse of strings.
This delocalization transition is qualitatively explained as follows.
In the case of the Heisenberg spin chain, the effect of interactions in scattering states is smaller than that in bound states \cite{Supple,karbach}.
The interactions of the ASEP are hardcore interactions. 
In the case of the TASEP, particles move in only one direction.
The collisions of particles are less likely to occur in the TASEP than in the ASEP ($p\neq1$), and the effect of the interactions is smaller.
Therefore, the string solutions collapse in the TASEP and bound states transition to scattering states.

\textit{Conclusion.}---
We have formulated the string solutions of the Bethe equations in the ASEP and elucidated the delocalization transition induced by the asymmetric hopping based on the picture of string-type quasiparticles.
The perspective of the asymmetry-induced transition by the collapse of strings would be applicable to a variety of situations since the strings are universal quasiparticles widely observed in quantum integrable systems.
Recently, related work was proposed where quasiparticles under dissipations are discussed in the context of quantum open systems \cite{haga2022quasiparticles}.
Understanding the quasiparticle picture in the general non-Hermitian condition is an important challenge for the future.

Integrable systems have celebrated analytical frameworks such as  the TBA \cite{takahashi_1999,yang1969thermodynamics,takahashi1971one,PhysRevLett.82.1764,mossel2012generalized} and the GHD \cite{PhysRevX.6.041065,PhysRevLett.117.207201,10.21468/SciPostPhys.6.4.049,10.21468/SciPostPhysLectNotes.18,RevModPhys.93.025003}.
These frameworks are not restricted to standard quantum systems. 
They are also applicable to classical integrable systems \cite{doyon2019generalized,Bonnemain_2022} and integrable cellular automata \cite{Kuniba_2020,croydon2021generalized} since the underlying quasiparticle picture is common in integrable systems.
By extending these methods to non-Hermitian systems, we would establish the analytical framework for the non-equilibrium systems, including stochastic processes and open quantum systems. Our results provide the first step toward this goal.

The authors thank Hosho Kastura for fruitful discussions.
This work was supported by JSPS KAKENHI Grant Number JP18K03448.






\bibliography{apssamp}

\providecommand{\noopsort}[1]{}\providecommand{\singleletter}[1]{#1}%
\begin{thebibliography}{55}%
\makeatletter
\providecommand \@ifxundefined [1]{%
 \@ifx{#1\undefined}
}%
\providecommand \@ifnum [1]{%
 \ifnum #1\expandafter \@firstoftwo
 \else \expandafter \@secondoftwo
 \fi
}%
\providecommand \@ifx [1]{%
 \ifx #1\expandafter \@firstoftwo
 \else \expandafter \@secondoftwo
 \fi
}%
\providecommand \natexlab [1]{#1}%
\providecommand \enquote  [1]{``#1''}%
\providecommand \bibnamefont  [1]{#1}%
\providecommand \bibfnamefont [1]{#1}%
\providecommand \citenamefont [1]{#1}%
\providecommand \href@noop [0]{\@secondoftwo}%
\providecommand \href [0]{\begingroup \@sanitize@url \@href}%
\providecommand \@href[1]{\@@startlink{#1}\@@href}%
\providecommand \@@href[1]{\endgroup#1\@@endlink}%
\providecommand \@sanitize@url [0]{\catcode `\\12\catcode `\$12\catcode
  `\&12\catcode `\#12\catcode `\^12\catcode `\_12\catcode `\%12\relax}%
\providecommand \@@startlink[1]{}%
\providecommand \@@endlink[0]{}%
\providecommand \url  [0]{\begingroup\@sanitize@url \@url }%
\providecommand \@url [1]{\endgroup\@href {#1}{\urlprefix }}%
\providecommand \urlprefix  [0]{URL }%
\providecommand \Eprint [0]{\href }%
\providecommand \doibase [0]{https://doi.org/}%
\providecommand \selectlanguage [0]{\@gobble}%
\providecommand \bibinfo  [0]{\@secondoftwo}%
\providecommand \bibfield  [0]{\@secondoftwo}%
\providecommand \translation [1]{[#1]}%
\providecommand \BibitemOpen [0]{}%
\providecommand \bibitemStop [0]{}%
\providecommand \bibitemNoStop [0]{.\EOS\space}%
\providecommand \EOS [0]{\spacefactor3000\relax}%
\providecommand \BibitemShut  [1]{\csname bibitem#1\endcsname}%
\let\auto@bib@innerbib\@empty
\bibitem [{\citenamefont {El-Ganainy}\ \emph {et~al.}(2018)\citenamefont
  {El-Ganainy}, \citenamefont {Makris}, \citenamefont {Khajavikhan},
  \citenamefont {Musslimani}, \citenamefont {Rotter},\ and\ \citenamefont
  {Christodoulides}}]{el2018non}%
  \BibitemOpen
  \bibfield  {author} {\bibinfo {author} {\bibfnamefont {R.}~\bibnamefont
  {El-Ganainy}}, \bibinfo {author} {\bibfnamefont {K.~G.}\ \bibnamefont
  {Makris}}, \bibinfo {author} {\bibfnamefont {M.}~\bibnamefont {Khajavikhan}},
  \bibinfo {author} {\bibfnamefont {Z.~H.}\ \bibnamefont {Musslimani}},
  \bibinfo {author} {\bibfnamefont {S.}~\bibnamefont {Rotter}},\ and\ \bibinfo
  {author} {\bibfnamefont {D.~N.}\ \bibnamefont {Christodoulides}},\
  }\href@noop {} {\bibfield  {journal} {\bibinfo  {journal} {Nat. Phys.}\
  }\textbf {\bibinfo {volume} {14}},\ \bibinfo {pages} {11} (\bibinfo {year}
  {2018})}\BibitemShut {NoStop}%
\bibitem [{\citenamefont {Ashida}\ \emph {et~al.}(2020)\citenamefont {Ashida},
  \citenamefont {Gong},\ and\ \citenamefont {Ueda}}]{ashida2020non}%
  \BibitemOpen
  \bibfield  {author} {\bibinfo {author} {\bibfnamefont {Y.}~\bibnamefont
  {Ashida}}, \bibinfo {author} {\bibfnamefont {Z.}~\bibnamefont {Gong}},\ and\
  \bibinfo {author} {\bibfnamefont {M.}~\bibnamefont {Ueda}},\ }\href@noop {}
  {\bibfield  {journal} {\bibinfo  {journal} {Adv. Phys.}\ }\textbf {\bibinfo
  {volume} {69}},\ \bibinfo {pages} {249} (\bibinfo {year} {2020})}\BibitemShut
  {NoStop}%
\bibitem [{\citenamefont {Derrida}(1998)}]{derrida1998exactly}%
  \BibitemOpen
  \bibfield  {author} {\bibinfo {author} {\bibfnamefont {B.}~\bibnamefont
  {Derrida}},\ }\href@noop {} {\bibfield  {journal} {\bibinfo  {journal} {Phys.
  Rep.}\ }\textbf {\bibinfo {volume} {301}},\ \bibinfo {pages} {65} (\bibinfo
  {year} {1998})}\BibitemShut {NoStop}%
\bibitem [{\citenamefont {Golinelli}\ and\ \citenamefont
  {Mallick}(2006)}]{golinelli2006asymmetric}%
  \BibitemOpen
  \bibfield  {author} {\bibinfo {author} {\bibfnamefont {O.}~\bibnamefont
  {Golinelli}}\ and\ \bibinfo {author} {\bibfnamefont {K.}~\bibnamefont
  {Mallick}},\ }\href@noop {} {\bibfield  {journal} {\bibinfo  {journal} {J.
  Phys. A: Math. Gen.}\ }\textbf {\bibinfo {volume} {39}},\ \bibinfo {pages}
  {12679} (\bibinfo {year} {2006})}\BibitemShut {NoStop}%
\bibitem [{\citenamefont {Blythe}\ and\ \citenamefont
  {Evans}(2007)}]{blythe2007nonequilibrium}%
  \BibitemOpen
  \bibfield  {author} {\bibinfo {author} {\bibfnamefont {R.~A.}\ \bibnamefont
  {Blythe}}\ and\ \bibinfo {author} {\bibfnamefont {M.~R.}\ \bibnamefont
  {Evans}},\ }\href@noop {} {\bibfield  {journal} {\bibinfo  {journal} {J.
  Phys. A: Math. Theor.}\ }\textbf {\bibinfo {volume} {40}},\ \bibinfo {pages}
  {R333} (\bibinfo {year} {2007})}\BibitemShut {NoStop}%
\bibitem [{\citenamefont {D\"urr}\ \emph {et~al.}(2009)\citenamefont {D\"urr},
  \citenamefont {Garc\'{\i}a-Ripoll}, \citenamefont {Syassen}, \citenamefont
  {Bauer}, \citenamefont {Lettner}, \citenamefont {Cirac},\ and\ \citenamefont
  {Rempe}}]{PhysRevA.79.023614}%
  \BibitemOpen
  \bibfield  {author} {\bibinfo {author} {\bibfnamefont {S.}~\bibnamefont
  {D\"urr}}, \bibinfo {author} {\bibfnamefont {J.~J.}\ \bibnamefont
  {Garc\'{\i}a-Ripoll}}, \bibinfo {author} {\bibfnamefont {N.}~\bibnamefont
  {Syassen}}, \bibinfo {author} {\bibfnamefont {D.~M.}\ \bibnamefont {Bauer}},
  \bibinfo {author} {\bibfnamefont {M.}~\bibnamefont {Lettner}}, \bibinfo
  {author} {\bibfnamefont {J.~I.}\ \bibnamefont {Cirac}},\ and\ \bibinfo
  {author} {\bibfnamefont {G.}~\bibnamefont {Rempe}},\ }\href
  {https://doi.org/10.1103/PhysRevA.79.023614} {\bibfield  {journal} {\bibinfo
  {journal} {Phys. Rev. A}\ }\textbf {\bibinfo {volume} {79}},\ \bibinfo
  {pages} {023614} (\bibinfo {year} {2009})}\BibitemShut {NoStop}%
\bibitem [{\citenamefont {Nakagawa}\ \emph {et~al.}(2021)\citenamefont
  {Nakagawa}, \citenamefont {Kawakami},\ and\ \citenamefont
  {Ueda}}]{PhysRevLett.126.110404}%
  \BibitemOpen
  \bibfield  {author} {\bibinfo {author} {\bibfnamefont {M.}~\bibnamefont
  {Nakagawa}}, \bibinfo {author} {\bibfnamefont {N.}~\bibnamefont {Kawakami}},\
  and\ \bibinfo {author} {\bibfnamefont {M.}~\bibnamefont {Ueda}},\ }\href
  {https://doi.org/10.1103/PhysRevLett.126.110404} {\bibfield  {journal}
  {\bibinfo  {journal} {Phys. Rev. Lett.}\ }\textbf {\bibinfo {volume} {126}},\
  \bibinfo {pages} {110404} (\bibinfo {year} {2021})}\BibitemShut {NoStop}%
\bibitem [{\citenamefont {Dalibard}\ \emph {et~al.}(1992)\citenamefont
  {Dalibard}, \citenamefont {Castin},\ and\ \citenamefont
  {M\o{}lmer}}]{PhysRevLett.68.580}%
  \BibitemOpen
  \bibfield  {author} {\bibinfo {author} {\bibfnamefont {J.}~\bibnamefont
  {Dalibard}}, \bibinfo {author} {\bibfnamefont {Y.}~\bibnamefont {Castin}},\
  and\ \bibinfo {author} {\bibfnamefont {K.}~\bibnamefont {M\o{}lmer}},\ }\href
  {https://doi.org/10.1103/PhysRevLett.68.580} {\bibfield  {journal} {\bibinfo
  {journal} {Phys. Rev. Lett.}\ }\textbf {\bibinfo {volume} {68}},\ \bibinfo
  {pages} {580} (\bibinfo {year} {1992})}\BibitemShut {NoStop}%
\bibitem [{\citenamefont {Korepin}\ \emph {et~al.}(1997)\citenamefont
  {Korepin}, \citenamefont {Bogoliubov},\ and\ \citenamefont
  {Izergin}}]{korepin1997quantum}%
  \BibitemOpen
  \bibfield  {author} {\bibinfo {author} {\bibfnamefont {V.~E.}\ \bibnamefont
  {Korepin}}, \bibinfo {author} {\bibfnamefont {N.~M.}\ \bibnamefont
  {Bogoliubov}},\ and\ \bibinfo {author} {\bibfnamefont {A.~G.}\ \bibnamefont
  {Izergin}},\ }\href@noop {} {\emph {\bibinfo {title} {Quantum inverse
  scattering method and correlation functions}}},\ Vol.~\bibinfo {volume} {3}\
  (\bibinfo  {publisher} {Cambridge university press},\ \bibinfo {year}
  {1997})\BibitemShut {NoStop}%
\bibitem [{\citenamefont {Takahashi}(1999)}]{takahashi_1999}%
  \BibitemOpen
  \bibfield  {author} {\bibinfo {author} {\bibfnamefont {M.}~\bibnamefont
  {Takahashi}},\ }\href {https://doi.org/10.1017/CBO9780511524332} {\emph
  {\bibinfo {title} {Thermodynamics of {O}ne-{D}imensional {S}olvable
  {M}odels}}}\ (\bibinfo  {publisher} {Cambridge University Press},\ \bibinfo
  {year} {1999})\BibitemShut {NoStop}%
\bibitem [{\citenamefont {Vlijm}\ \emph {et~al.}(2015)\citenamefont {Vlijm},
  \citenamefont {Ganahl}, \citenamefont {Fioretto}, \citenamefont {Brockmann},
  \citenamefont {Haque}, \citenamefont {Evertz},\ and\ \citenamefont
  {Caux}}]{PhysRevB.92.214427}%
  \BibitemOpen
  \bibfield  {author} {\bibinfo {author} {\bibfnamefont {R.}~\bibnamefont
  {Vlijm}}, \bibinfo {author} {\bibfnamefont {M.}~\bibnamefont {Ganahl}},
  \bibinfo {author} {\bibfnamefont {D.}~\bibnamefont {Fioretto}}, \bibinfo
  {author} {\bibfnamefont {M.}~\bibnamefont {Brockmann}}, \bibinfo {author}
  {\bibfnamefont {M.}~\bibnamefont {Haque}}, \bibinfo {author} {\bibfnamefont
  {H.~G.}\ \bibnamefont {Evertz}},\ and\ \bibinfo {author} {\bibfnamefont
  {J.-S.}\ \bibnamefont {Caux}},\ }\href
  {https://doi.org/10.1103/PhysRevB.92.214427} {\bibfield  {journal} {\bibinfo
  {journal} {Phys. Rev. B}\ }\textbf {\bibinfo {volume} {92}},\ \bibinfo
  {pages} {214427} (\bibinfo {year} {2015})}\BibitemShut {NoStop}%
\bibitem [{\citenamefont {Wadachi}\ and\ \citenamefont
  {Sakagami}(1984)}]{wadachi1984classical}%
  \BibitemOpen
  \bibfield  {author} {\bibinfo {author} {\bibfnamefont {M.}~\bibnamefont
  {Wadachi}}\ and\ \bibinfo {author} {\bibfnamefont {M.}~\bibnamefont
  {Sakagami}},\ }\href@noop {} {\bibfield  {journal} {\bibinfo  {journal} {J.
  Phys. Soc. Jpn.}\ }\textbf {\bibinfo {volume} {53}},\ \bibinfo {pages} {1933}
  (\bibinfo {year} {1984})}\BibitemShut {NoStop}%
\bibitem [{\citenamefont {Wadati}\ \emph {et~al.}(1985)\citenamefont {Wadati},
  \citenamefont {Kuniba},\ and\ \citenamefont {Konishi}}]{wadati1985quantum}%
  \BibitemOpen
  \bibfield  {author} {\bibinfo {author} {\bibfnamefont {M.}~\bibnamefont
  {Wadati}}, \bibinfo {author} {\bibfnamefont {A.}~\bibnamefont {Kuniba}},\
  and\ \bibinfo {author} {\bibfnamefont {T.}~\bibnamefont {Konishi}},\
  }\href@noop {} {\bibfield  {journal} {\bibinfo  {journal} {J. Phys. Soc.
  Jpn.}\ }\textbf {\bibinfo {volume} {54}},\ \bibinfo {pages} {1710} (\bibinfo
  {year} {1985})}\BibitemShut {NoStop}%
\bibitem [{\citenamefont {Lai}\ and\ \citenamefont
  {Haus}(1989)}]{PhysRevA.40.854}%
  \BibitemOpen
  \bibfield  {author} {\bibinfo {author} {\bibfnamefont {Y.}~\bibnamefont
  {Lai}}\ and\ \bibinfo {author} {\bibfnamefont {H.~A.}\ \bibnamefont {Haus}},\
  }\href {https://doi.org/10.1103/PhysRevA.40.854} {\bibfield  {journal}
  {\bibinfo  {journal} {Phys. Rev. A}\ }\textbf {\bibinfo {volume} {40}},\
  \bibinfo {pages} {854} (\bibinfo {year} {1989})}\BibitemShut {NoStop}%
\bibitem [{\citenamefont {De~Nardis}\ \emph {et~al.}(2020)\citenamefont
  {De~Nardis}, \citenamefont {Gopalakrishnan}, \citenamefont {Ilievski},\ and\
  \citenamefont {Vasseur}}]{PhysRevLett.125.070601}%
  \BibitemOpen
  \bibfield  {author} {\bibinfo {author} {\bibfnamefont {J.}~\bibnamefont
  {De~Nardis}}, \bibinfo {author} {\bibfnamefont {S.}~\bibnamefont
  {Gopalakrishnan}}, \bibinfo {author} {\bibfnamefont {E.}~\bibnamefont
  {Ilievski}},\ and\ \bibinfo {author} {\bibfnamefont {R.}~\bibnamefont
  {Vasseur}},\ }\href {https://doi.org/10.1103/PhysRevLett.125.070601}
  {\bibfield  {journal} {\bibinfo  {journal} {Phys. Rev. Lett.}\ }\textbf
  {\bibinfo {volume} {125}},\ \bibinfo {pages} {070601} (\bibinfo {year}
  {2020})}\BibitemShut {NoStop}%
\bibitem [{\citenamefont {Bulchandani}\ \emph {et~al.}(2021)\citenamefont
  {Bulchandani}, \citenamefont {Gopalakrishnan},\ and\ \citenamefont
  {Ilievski}}]{Bulchandani_2021}%
  \BibitemOpen
  \bibfield  {author} {\bibinfo {author} {\bibfnamefont {V.~B.}\ \bibnamefont
  {Bulchandani}}, \bibinfo {author} {\bibfnamefont {S.}~\bibnamefont
  {Gopalakrishnan}},\ and\ \bibinfo {author} {\bibfnamefont {E.}~\bibnamefont
  {Ilievski}},\ }\href {https://doi.org/10.1088/1742-5468/ac12c7} {\bibfield
  {journal} {\bibinfo  {journal} {Journal of Statistical Mechanics: Theory and
  Experiment}\ }\textbf {\bibinfo {volume} {2021}},\ \bibinfo {pages} {084001}
  (\bibinfo {year} {2021})}\BibitemShut {NoStop}%
\bibitem [{\citenamefont {Yang}\ and\ \citenamefont
  {Yang}(1969)}]{yang1969thermodynamics}%
  \BibitemOpen
  \bibfield  {author} {\bibinfo {author} {\bibfnamefont {C.-N.}\ \bibnamefont
  {Yang}}\ and\ \bibinfo {author} {\bibfnamefont {C.~P.}\ \bibnamefont
  {Yang}},\ }\href@noop {} {\bibfield  {journal} {\bibinfo  {journal} {J. Math.
  Phys.}\ }\textbf {\bibinfo {volume} {10}},\ \bibinfo {pages} {1115} (\bibinfo
  {year} {1969})}\BibitemShut {NoStop}%
\bibitem [{\citenamefont {Takahashi}(1971)}]{takahashi1971one}%
  \BibitemOpen
  \bibfield  {author} {\bibinfo {author} {\bibfnamefont {M.}~\bibnamefont
  {Takahashi}},\ }\href@noop {} {\bibfield  {journal} {\bibinfo  {journal}
  {Prog. Theor. Phys.}\ }\textbf {\bibinfo {volume} {46}},\ \bibinfo {pages}
  {401} (\bibinfo {year} {1971})}\BibitemShut {NoStop}%
\bibitem [{\citenamefont {Zotos}(1999)}]{PhysRevLett.82.1764}%
  \BibitemOpen
  \bibfield  {author} {\bibinfo {author} {\bibfnamefont {X.}~\bibnamefont
  {Zotos}},\ }\href {https://doi.org/10.1103/PhysRevLett.82.1764} {\bibfield
  {journal} {\bibinfo  {journal} {Phys. Rev. Lett.}\ }\textbf {\bibinfo
  {volume} {82}},\ \bibinfo {pages} {1764} (\bibinfo {year}
  {1999})}\BibitemShut {NoStop}%
\bibitem [{\citenamefont {Mossel}\ and\ \citenamefont
  {Caux}(2012)}]{mossel2012generalized}%
  \BibitemOpen
  \bibfield  {author} {\bibinfo {author} {\bibfnamefont {J.}~\bibnamefont
  {Mossel}}\ and\ \bibinfo {author} {\bibfnamefont {J.-S.}\ \bibnamefont
  {Caux}},\ }\href@noop {} {\bibfield  {journal} {\bibinfo  {journal} {J. Phys.
  A: Math. Theor.}\ }\textbf {\bibinfo {volume} {45}},\ \bibinfo {pages}
  {255001} (\bibinfo {year} {2012})}\BibitemShut {NoStop}%
\bibitem [{\citenamefont {Castro-Alvaredo}\ \emph {et~al.}(2016)\citenamefont
  {Castro-Alvaredo}, \citenamefont {Doyon},\ and\ \citenamefont
  {Yoshimura}}]{PhysRevX.6.041065}%
  \BibitemOpen
  \bibfield  {author} {\bibinfo {author} {\bibfnamefont {O.~A.}\ \bibnamefont
  {Castro-Alvaredo}}, \bibinfo {author} {\bibfnamefont {B.}~\bibnamefont
  {Doyon}},\ and\ \bibinfo {author} {\bibfnamefont {T.}~\bibnamefont
  {Yoshimura}},\ }\href {https://doi.org/10.1103/PhysRevX.6.041065} {\bibfield
  {journal} {\bibinfo  {journal} {Phys. Rev. X}\ }\textbf {\bibinfo {volume}
  {6}},\ \bibinfo {pages} {041065} (\bibinfo {year} {2016})}\BibitemShut
  {NoStop}%
\bibitem [{\citenamefont {Bertini}\ \emph {et~al.}(2016)\citenamefont
  {Bertini}, \citenamefont {Collura}, \citenamefont {De~Nardis},\ and\
  \citenamefont {Fagotti}}]{PhysRevLett.117.207201}%
  \BibitemOpen
  \bibfield  {author} {\bibinfo {author} {\bibfnamefont {B.}~\bibnamefont
  {Bertini}}, \bibinfo {author} {\bibfnamefont {M.}~\bibnamefont {Collura}},
  \bibinfo {author} {\bibfnamefont {J.}~\bibnamefont {De~Nardis}},\ and\
  \bibinfo {author} {\bibfnamefont {M.}~\bibnamefont {Fagotti}},\ }\href
  {https://doi.org/10.1103/PhysRevLett.117.207201} {\bibfield  {journal}
  {\bibinfo  {journal} {Phys. Rev. Lett.}\ }\textbf {\bibinfo {volume} {117}},\
  \bibinfo {pages} {207201} (\bibinfo {year} {2016})}\BibitemShut {NoStop}%
\bibitem [{\citenamefont {Nardis}\ \emph {et~al.}(2019)\citenamefont {Nardis},
  \citenamefont {Bernard},\ and\ \citenamefont
  {Doyon}}]{10.21468/SciPostPhys.6.4.049}%
  \BibitemOpen
  \bibfield  {author} {\bibinfo {author} {\bibfnamefont {J.~D.}\ \bibnamefont
  {Nardis}}, \bibinfo {author} {\bibfnamefont {D.}~\bibnamefont {Bernard}},\
  and\ \bibinfo {author} {\bibfnamefont {B.}~\bibnamefont {Doyon}},\ }\href
  {https://doi.org/10.21468/SciPostPhys.6.4.049} {\bibfield  {journal}
  {\bibinfo  {journal} {SciPost Phys.}\ }\textbf {\bibinfo {volume} {6}},\
  \bibinfo {pages} {049} (\bibinfo {year} {2019})}\BibitemShut {NoStop}%
\bibitem [{\citenamefont {Doyon}(2020)}]{10.21468/SciPostPhysLectNotes.18}%
  \BibitemOpen
  \bibfield  {author} {\bibinfo {author} {\bibfnamefont {B.}~\bibnamefont
  {Doyon}},\ }\href {https://doi.org/10.21468/SciPostPhysLectNotes.18}
  {\bibfield  {journal} {\bibinfo  {journal} {SciPost Phys. Lect. Notes}\ ,\
  \bibinfo {pages} {18}} (\bibinfo {year} {2020})}\BibitemShut {NoStop}%
\bibitem [{\citenamefont {Bertini}\ \emph {et~al.}(2021)\citenamefont
  {Bertini}, \citenamefont {Heidrich-Meisner}, \citenamefont {Karrasch},
  \citenamefont {Prosen}, \citenamefont {Steinigeweg},\ and\ \citenamefont
  {\ifmmode \check{Z}\else \v{Z}\fi{}nidari\ifmmode~\check{c}\else
  \v{c}\fi{}}}]{RevModPhys.93.025003}%
  \BibitemOpen
  \bibfield  {author} {\bibinfo {author} {\bibfnamefont {B.}~\bibnamefont
  {Bertini}}, \bibinfo {author} {\bibfnamefont {F.}~\bibnamefont
  {Heidrich-Meisner}}, \bibinfo {author} {\bibfnamefont {C.}~\bibnamefont
  {Karrasch}}, \bibinfo {author} {\bibfnamefont {T.}~\bibnamefont {Prosen}},
  \bibinfo {author} {\bibfnamefont {R.}~\bibnamefont {Steinigeweg}},\ and\
  \bibinfo {author} {\bibfnamefont {M.}~\bibnamefont {\ifmmode \check{Z}\else
  \v{Z}\fi{}nidari\ifmmode~\check{c}\else \v{c}\fi{}}},\ }\href
  {https://doi.org/10.1103/RevModPhys.93.025003} {\bibfield  {journal}
  {\bibinfo  {journal} {Rev. Mod. Phys.}\ }\textbf {\bibinfo {volume} {93}},\
  \bibinfo {pages} {025003} (\bibinfo {year} {2021})}\BibitemShut {NoStop}%
\bibitem [{\citenamefont {Essler}\ and\ \citenamefont
  {Rittenberg}(1996)}]{essler1996representations}%
  \BibitemOpen
  \bibfield  {author} {\bibinfo {author} {\bibfnamefont {F.~H.}\ \bibnamefont
  {Essler}}\ and\ \bibinfo {author} {\bibfnamefont {V.}~\bibnamefont
  {Rittenberg}},\ }\href@noop {} {\bibfield  {journal} {\bibinfo  {journal} {J.
  Phys. A: Math. Gen.}\ }\textbf {\bibinfo {volume} {29}},\ \bibinfo {pages}
  {3375} (\bibinfo {year} {1996})}\BibitemShut {NoStop}%
\bibitem [{\citenamefont {Gwa}\ and\ \citenamefont
  {Spohn}(1992)}]{gwa1992bethe}%
  \BibitemOpen
  \bibfield  {author} {\bibinfo {author} {\bibfnamefont {L.-H.}\ \bibnamefont
  {Gwa}}\ and\ \bibinfo {author} {\bibfnamefont {H.}~\bibnamefont {Spohn}},\
  }\href@noop {} {\bibfield  {journal} {\bibinfo  {journal} {Phys. Rev. A}\
  }\textbf {\bibinfo {volume} {46}},\ \bibinfo {pages} {844} (\bibinfo {year}
  {1992})}\BibitemShut {NoStop}%
\bibitem [{\citenamefont {Kim}(1995)}]{kim1995bethe}%
  \BibitemOpen
  \bibfield  {author} {\bibinfo {author} {\bibfnamefont {D.}~\bibnamefont
  {Kim}},\ }\href@noop {} {\bibfield  {journal} {\bibinfo  {journal} {Phys.
  Rev. E}\ }\textbf {\bibinfo {volume} {52}},\ \bibinfo {pages} {3512}
  (\bibinfo {year} {1995})}\BibitemShut {NoStop}%
\bibitem [{\citenamefont {Golinelli}\ and\ \citenamefont
  {Mallick}(2004)}]{Golinelli_2004}%
  \BibitemOpen
  \bibfield  {author} {\bibinfo {author} {\bibfnamefont {O.}~\bibnamefont
  {Golinelli}}\ and\ \bibinfo {author} {\bibfnamefont {K.}~\bibnamefont
  {Mallick}},\ }\href {https://doi.org/10.1088/0305-4470/37/10/001} {\bibfield
  {journal} {\bibinfo  {journal} {J. Phys. A: Math. Gen.}\ }\textbf {\bibinfo
  {volume} {37}},\ \bibinfo {pages} {3321} (\bibinfo {year}
  {2004})}\BibitemShut {NoStop}%
\bibitem [{\citenamefont {Golinelli}\ and\ \citenamefont
  {Mallick}(2005)}]{Golinelli_2005}%
  \BibitemOpen
  \bibfield  {author} {\bibinfo {author} {\bibfnamefont {O.}~\bibnamefont
  {Golinelli}}\ and\ \bibinfo {author} {\bibfnamefont {K.}~\bibnamefont
  {Mallick}},\ }\href {https://doi.org/10.1088/0305-4470/38/7/001} {\bibfield
  {journal} {\bibinfo  {journal} {J. Phys. A: Math. Gen.}\ }\textbf {\bibinfo
  {volume} {38}},\ \bibinfo {pages} {1419} (\bibinfo {year}
  {2005})}\BibitemShut {NoStop}%
\bibitem [{\citenamefont {Prolhac}(2013)}]{prolhac2013spectrum}%
  \BibitemOpen
  \bibfield  {author} {\bibinfo {author} {\bibfnamefont {S.}~\bibnamefont
  {Prolhac}},\ }\href@noop {} {\bibfield  {journal} {\bibinfo  {journal} {J.
  Phys. A: Math. Theor.}\ }\textbf {\bibinfo {volume} {46}},\ \bibinfo {pages}
  {415001} (\bibinfo {year} {2013})}\BibitemShut {NoStop}%
\bibitem [{\citenamefont {Prolhac}(2014)}]{prolhac2014spectrum}%
  \BibitemOpen
  \bibfield  {author} {\bibinfo {author} {\bibfnamefont {S.}~\bibnamefont
  {Prolhac}},\ }\href@noop {} {\bibfield  {journal} {\bibinfo  {journal} {J.
  Phys. A: Math. Theor.}\ }\textbf {\bibinfo {volume} {47}},\ \bibinfo {pages}
  {375001} (\bibinfo {year} {2014})}\BibitemShut {NoStop}%
\bibitem [{\citenamefont {Prolhac}(2016)}]{prolhac2016extrapolation}%
  \BibitemOpen
  \bibfield  {author} {\bibinfo {author} {\bibfnamefont {S.}~\bibnamefont
  {Prolhac}},\ }\href@noop {} {\bibfield  {journal} {\bibinfo  {journal} {J.
  Phys. A: Math. Theor.}\ }\textbf {\bibinfo {volume} {49}},\ \bibinfo {pages}
  {454002} (\bibinfo {year} {2016})}\BibitemShut {NoStop}%
\bibitem [{\citenamefont {Prolhac}(2017)}]{prolhac2017perturbative}%
  \BibitemOpen
  \bibfield  {author} {\bibinfo {author} {\bibfnamefont {S.}~\bibnamefont
  {Prolhac}},\ }\href@noop {} {\bibfield  {journal} {\bibinfo  {journal} {J.
  Phys. A: Math. Theor.}\ }\textbf {\bibinfo {volume} {50}},\ \bibinfo {pages}
  {315001} (\bibinfo {year} {2017})}\BibitemShut {NoStop}%
\bibitem [{\citenamefont {De~Gier}\ and\ \citenamefont
  {Essler}(2005)}]{de2005bethe}%
  \BibitemOpen
  \bibfield  {author} {\bibinfo {author} {\bibfnamefont {J.}~\bibnamefont
  {De~Gier}}\ and\ \bibinfo {author} {\bibfnamefont {F.~H.}\ \bibnamefont
  {Essler}},\ }\href@noop {} {\bibfield  {journal} {\bibinfo  {journal} {Phys.
  Rev. Lett.}\ }\textbf {\bibinfo {volume} {95}},\ \bibinfo {pages} {240601}
  (\bibinfo {year} {2005})}\BibitemShut {NoStop}%
\bibitem [{\citenamefont {de~Gier}\ and\ \citenamefont
  {Essler}(2006)}]{deGier_2006}%
  \BibitemOpen
  \bibfield  {author} {\bibinfo {author} {\bibfnamefont {J.}~\bibnamefont
  {de~Gier}}\ and\ \bibinfo {author} {\bibfnamefont {F.~H.~L.}\ \bibnamefont
  {Essler}},\ }\href {https://doi.org/10.1088/1742-5468/2006/12/P12011}
  {\bibfield  {journal} {\bibinfo  {journal} {J. Stat. Mech.: Theory Exp.}\
  }\textbf {\bibinfo {volume} {2006}}\bibinfo  {number} { (12)},\ \bibinfo
  {pages} {P12011}}\BibitemShut {NoStop}%
\bibitem [{\citenamefont {de~Gier}\ and\ \citenamefont
  {Essler}(2008)}]{deGier_2008}%
  \BibitemOpen
\bibfield  {number} {  }\bibfield  {author} {\bibinfo {author} {\bibfnamefont
  {J.}~\bibnamefont {de~Gier}}\ and\ \bibinfo {author} {\bibfnamefont
  {F.~H.~L.}\ \bibnamefont {Essler}},\ }\href
  {https://doi.org/10.1088/1751-8113/41/48/485002} {\bibfield  {journal}
  {\bibinfo  {journal} {J. Phys. A: Math. Theor.}\ }\textbf {\bibinfo {volume}
  {41}},\ \bibinfo {pages} {485002} (\bibinfo {year} {2008})}\BibitemShut
  {NoStop}%
\bibitem [{\citenamefont {de~Gier}\ and\ \citenamefont
  {Essler}(2011)}]{PhysRevLett.107.010602}%
  \BibitemOpen
  \bibfield  {author} {\bibinfo {author} {\bibfnamefont {J.}~\bibnamefont
  {de~Gier}}\ and\ \bibinfo {author} {\bibfnamefont {F.~H.~L.}\ \bibnamefont
  {Essler}},\ }\href {https://doi.org/10.1103/PhysRevLett.107.010602}
  {\bibfield  {journal} {\bibinfo  {journal} {Phys. Rev. Lett.}\ }\textbf
  {\bibinfo {volume} {107}},\ \bibinfo {pages} {010602} (\bibinfo {year}
  {2011})}\BibitemShut {NoStop}%
\bibitem [{\citenamefont {Wen}\ \emph {et~al.}(2015)\citenamefont {Wen},
  \citenamefont {Yang}, \citenamefont {Cui}, \citenamefont {Cao},\ and\
  \citenamefont {Yang}}]{Wen_2015}%
  \BibitemOpen
  \bibfield  {author} {\bibinfo {author} {\bibfnamefont {F.-K.}\ \bibnamefont
  {Wen}}, \bibinfo {author} {\bibfnamefont {Z.-Y.}\ \bibnamefont {Yang}},
  \bibinfo {author} {\bibfnamefont {S.}~\bibnamefont {Cui}}, \bibinfo {author}
  {\bibfnamefont {J.-P.}\ \bibnamefont {Cao}},\ and\ \bibinfo {author}
  {\bibfnamefont {W.-L.}\ \bibnamefont {Yang}},\ }\href
  {https://doi.org/10.1088/0256-307X/32/5/050503} {\bibfield  {journal}
  {\bibinfo  {journal} {Chin. Phys. Lett.}\ }\textbf {\bibinfo {volume} {32}},\
  \bibinfo {pages} {050503} (\bibinfo {year} {2015})}\BibitemShut {NoStop}%
\bibitem [{\citenamefont {Crampé}(2015)}]{Crampe_2015}%
  \BibitemOpen
  \bibfield  {author} {\bibinfo {author} {\bibfnamefont {N.}~\bibnamefont
  {Crampé}},\ }\href {https://doi.org/10.1088/1751-8113/48/8/08FT01}
  {\bibfield  {journal} {\bibinfo  {journal} {J. Phys. A: Math. Theor.}\
  }\textbf {\bibinfo {volume} {48}},\ \bibinfo {pages} {08FT01} (\bibinfo
  {year} {2015})}\BibitemShut {NoStop}%
\bibitem [{\citenamefont {Bertini}\ and\ \citenamefont
  {Giacomin}(1997)}]{bertini1997stochastic}%
  \BibitemOpen
  \bibfield  {author} {\bibinfo {author} {\bibfnamefont {L.}~\bibnamefont
  {Bertini}}\ and\ \bibinfo {author} {\bibfnamefont {G.}~\bibnamefont
  {Giacomin}},\ }\href@noop {} {\bibfield  {journal} {\bibinfo  {journal}
  {Commun. Math. Phys.}\ }\textbf {\bibinfo {volume} {183}},\ \bibinfo {pages}
  {571} (\bibinfo {year} {1997})}\BibitemShut {NoStop}%
\bibitem [{\citenamefont {Takeuchi}(2018)}]{takeuchi2018appetizer}%
  \BibitemOpen
  \bibfield  {author} {\bibinfo {author} {\bibfnamefont {K.~A.}\ \bibnamefont
  {Takeuchi}},\ }\href@noop {} {\bibfield  {journal} {\bibinfo  {journal}
  {Physica A}\ }\textbf {\bibinfo {volume} {504}},\ \bibinfo {pages} {77}
  (\bibinfo {year} {2018})}\BibitemShut {NoStop}%
\bibitem [{\citenamefont
  {Schadschneider}(2000)}]{schadschneider2000statistical}%
  \BibitemOpen
  \bibfield  {author} {\bibinfo {author} {\bibfnamefont {A.}~\bibnamefont
  {Schadschneider}},\ }\href@noop {} {\bibfield  {journal} {\bibinfo  {journal}
  {Physica A}\ }\textbf {\bibinfo {volume} {285}},\ \bibinfo {pages} {101}
  (\bibinfo {year} {2000})}\BibitemShut {NoStop}%
\bibitem [{\citenamefont {Schadschneider}\ \emph {et~al.}(2010)\citenamefont
  {Schadschneider}, \citenamefont {Chowdhury},\ and\ \citenamefont
  {Nishinari}}]{schadschneider2010stochastic}%
  \BibitemOpen
  \bibfield  {author} {\bibinfo {author} {\bibfnamefont {A.}~\bibnamefont
  {Schadschneider}}, \bibinfo {author} {\bibfnamefont {D.}~\bibnamefont
  {Chowdhury}},\ and\ \bibinfo {author} {\bibfnamefont {K.}~\bibnamefont
  {Nishinari}},\ }\href@noop {} {\emph {\bibinfo {title} {Stochastic transport
  in complex systems: from molecules to vehicles}}}\ (\bibinfo  {publisher}
  {Elsevier},\ \bibinfo {year} {2010})\BibitemShut {NoStop}%
\bibitem [{\citenamefont {MacDonald}\ \emph {et~al.}(1968)\citenamefont
  {MacDonald}, \citenamefont {Gibbs},\ and\ \citenamefont
  {Pipkin}}]{macdonald1968kinetics}%
  \BibitemOpen
  \bibfield  {author} {\bibinfo {author} {\bibfnamefont {C.~T.}\ \bibnamefont
  {MacDonald}}, \bibinfo {author} {\bibfnamefont {J.~H.}\ \bibnamefont
  {Gibbs}},\ and\ \bibinfo {author} {\bibfnamefont {A.~C.}\ \bibnamefont
  {Pipkin}},\ }\href@noop {} {\bibfield  {journal} {\bibinfo  {journal}
  {Biopolymers}\ }\textbf {\bibinfo {volume} {6}},\ \bibinfo {pages} {1}
  (\bibinfo {year} {1968})}\BibitemShut {NoStop}%
\bibitem [{\citenamefont {Klumpp}\ and\ \citenamefont
  {Lipowsky}(2003)}]{klumpp2003traffic}%
  \BibitemOpen
  \bibfield  {author} {\bibinfo {author} {\bibfnamefont {S.}~\bibnamefont
  {Klumpp}}\ and\ \bibinfo {author} {\bibfnamefont {R.}~\bibnamefont
  {Lipowsky}},\ }\href@noop {} {\bibfield  {journal} {\bibinfo  {journal} {J.
  Stat. Phys.}\ }\textbf {\bibinfo {volume} {113}},\ \bibinfo {pages} {233}
  (\bibinfo {year} {2003})}\BibitemShut {NoStop}%
\bibitem [{Sup()}]{Supple}%
  \BibitemOpen
  \href@noop {} {}\bibinfo {note} {See Supplemental Material for the
  delocalization of Bethe quantum numbers, the detailed data of the numerical
  solutions, the complexification of string centers, the derivation of the
  Bethe-Takahashi equations, the proof of Eqs. (\ref{eq:lhscases}) and
  (\ref{eq:BTlhscases}), and the dispersion relation of the Heisenberg spin
  chain.}\BibitemShut {Stop}%
\bibitem [{\citenamefont {Vladimirov}(1986)}]{vladimirov1986proof}%
  \BibitemOpen
  \bibfield  {author} {\bibinfo {author} {\bibfnamefont {A.~A.}\ \bibnamefont
  {Vladimirov}},\ }\href {https://doi.org/10.1103/PhysRevA.79.023614}
  {\bibfield  {journal} {\bibinfo  {journal} {Ther. Math. Phys.}\ }\textbf
  {\bibinfo {volume} {66}},\ \bibinfo {pages} {102} (\bibinfo {year}
  {1986})}\BibitemShut {NoStop}%
\bibitem [{\citenamefont {Hagemans}\ and\ \citenamefont
  {Caux}(2007)}]{Hagemans_2007}%
  \BibitemOpen
  \bibfield  {author} {\bibinfo {author} {\bibfnamefont {R.}~\bibnamefont
  {Hagemans}}\ and\ \bibinfo {author} {\bibfnamefont {J.-S.}\ \bibnamefont
  {Caux}},\ }\href {https://doi.org/10.1088/1751-8113/40/49/001} {\bibfield
  {journal} {\bibinfo  {journal} {J. Phys. A: Math. Theor.}\ }\textbf {\bibinfo
  {volume} {40}},\ \bibinfo {pages} {14605} (\bibinfo {year}
  {2007})}\BibitemShut {NoStop}%
\bibitem [{\citenamefont {Karbach}\ and\ \citenamefont
  {Müller}(1997)}]{karbach}%
  \BibitemOpen
  \bibfield  {author} {\bibinfo {author} {\bibfnamefont {M.}~\bibnamefont
  {Karbach}}\ and\ \bibinfo {author} {\bibfnamefont {G.}~\bibnamefont
  {Müller}},\ }\href {https://doi.org/10.1063/1.4822511} {\bibfield  {journal}
  {\bibinfo  {journal} {Comput. phys.}\ }\textbf {\bibinfo {volume} {11}},\
  \bibinfo {pages} {36} (\bibinfo {year} {1997})}\BibitemShut {NoStop}%
\bibitem [{\citenamefont {Haga}\ \emph {et~al.}(2022)\citenamefont {Haga},
  \citenamefont {Nakagawa}, \citenamefont {Hamazaki},\ and\ \citenamefont
  {Ueda}}]{haga2022quasiparticles}%
  \BibitemOpen
  \bibfield  {author} {\bibinfo {author} {\bibfnamefont {T.}~\bibnamefont
  {Haga}}, \bibinfo {author} {\bibfnamefont {M.}~\bibnamefont {Nakagawa}},
  \bibinfo {author} {\bibfnamefont {R.}~\bibnamefont {Hamazaki}},\ and\
  \bibinfo {author} {\bibfnamefont {M.}~\bibnamefont {Ueda}},\ }\href@noop {}
  {\bibfield  {journal} {\bibinfo  {journal} {arXiv preprint arXiv:2211.14991}\
  } (\bibinfo {year} {2022})}\BibitemShut {NoStop}%
\bibitem [{\citenamefont {Doyon}(2019)}]{doyon2019generalized}%
  \BibitemOpen
  \bibfield  {author} {\bibinfo {author} {\bibfnamefont {B.}~\bibnamefont
  {Doyon}},\ }\href@noop {} {\bibfield  {journal} {\bibinfo  {journal} {Journal
  of Mathematical Physics}\ }\textbf {\bibinfo {volume} {60}},\ \bibinfo
  {pages} {073302} (\bibinfo {year} {2019})}\BibitemShut {NoStop}%
\bibitem [{\citenamefont {Bonnemain}\ \emph {et~al.}(2022)\citenamefont
  {Bonnemain}, \citenamefont {Doyon},\ and\ \citenamefont
  {El}}]{Bonnemain_2022}%
  \BibitemOpen
  \bibfield  {author} {\bibinfo {author} {\bibfnamefont {T.}~\bibnamefont
  {Bonnemain}}, \bibinfo {author} {\bibfnamefont {B.}~\bibnamefont {Doyon}},\
  and\ \bibinfo {author} {\bibfnamefont {G.}~\bibnamefont {El}},\ }\href
  {https://doi.org/10.1088/1751-8121/ac8253} {\bibfield  {journal} {\bibinfo
  {journal} {Journal of Physics A: Mathematical and Theoretical}\ }\textbf
  {\bibinfo {volume} {55}},\ \bibinfo {pages} {374004} (\bibinfo {year}
  {2022})}\BibitemShut {NoStop}%
\bibitem [{\citenamefont {Kuniba}\ \emph {et~al.}(2020)\citenamefont {Kuniba},
  \citenamefont {Misguich},\ and\ \citenamefont {Pasquier}}]{Kuniba_2020}%
  \BibitemOpen
  \bibfield  {author} {\bibinfo {author} {\bibfnamefont {A.}~\bibnamefont
  {Kuniba}}, \bibinfo {author} {\bibfnamefont {G.}~\bibnamefont {Misguich}},\
  and\ \bibinfo {author} {\bibfnamefont {V.}~\bibnamefont {Pasquier}},\ }\href
  {https://doi.org/10.1088/1751-8121/abadb9} {\bibfield  {journal} {\bibinfo
  {journal} {Journal of Physics A: Mathematical and Theoretical}\ }\textbf
  {\bibinfo {volume} {53}},\ \bibinfo {pages} {404001} (\bibinfo {year}
  {2020})}\BibitemShut {NoStop}%
\bibitem [{\citenamefont {Croydon}\ and\ \citenamefont
  {Sasada}(2021)}]{croydon2021generalized}%
  \BibitemOpen
  \bibfield  {author} {\bibinfo {author} {\bibfnamefont {D.~A.}\ \bibnamefont
  {Croydon}}\ and\ \bibinfo {author} {\bibfnamefont {M.}~\bibnamefont
  {Sasada}},\ }\href@noop {} {\bibfield  {journal} {\bibinfo  {journal}
  {Communications in Mathematical Physics}\ }\textbf {\bibinfo {volume}
  {383}},\ \bibinfo {pages} {427} (\bibinfo {year} {2021})}\BibitemShut
  {NoStop}%
\end{thebibliography}%


\providecommand{\noopsort}[1]{}\providecommand{\singleletter}[1]{#1}%
\begin{thebibliography}{3}%
\makeatletter
\providecommand \@ifxundefined [1]{%
 \@ifx{#1\undefined}
}%
\providecommand \@ifnum [1]{%
 \ifnum #1\expandafter \@firstoftwo
 \else \expandafter \@secondoftwo
 \fi
}%
\providecommand \@ifx [1]{%
 \ifx #1\expandafter \@firstoftwo
 \else \expandafter \@secondoftwo
 \fi
}%
\providecommand \natexlab [1]{#1}%
\providecommand \enquote  [1]{``#1''}%
\providecommand \bibnamefont  [1]{#1}%
\providecommand \bibfnamefont [1]{#1}%
\providecommand \citenamefont [1]{#1}%
\providecommand \href@noop [0]{\@secondoftwo}%
\providecommand \href [0]{\begingroup \@sanitize@url \@href}%
\providecommand \@href[1]{\@@startlink{#1}\@@href}%
\providecommand \@@href[1]{\endgroup#1\@@endlink}%
\providecommand \@sanitize@url [0]{\catcode `\\12\catcode `\$12\catcode
  `\&12\catcode `\#12\catcode `\^12\catcode `\_12\catcode `\%12\relax}%
\providecommand \@@startlink[1]{}%
\providecommand \@@endlink[0]{}%
\providecommand \url  [0]{\begingroup\@sanitize@url \@url }%
\providecommand \@url [1]{\endgroup\@href {#1}{\urlprefix }}%
\providecommand \urlprefix  [0]{URL }%
\providecommand \Eprint [0]{\href }%
\providecommand \doibase [0]{https://doi.org/}%
\providecommand \selectlanguage [0]{\@gobble}%
\providecommand \bibinfo  [0]{\@secondoftwo}%
\providecommand \bibfield  [0]{\@secondoftwo}%
\providecommand \translation [1]{[#1]}%
\providecommand \BibitemOpen [0]{}%
\providecommand \bibitemStop [0]{}%
\providecommand \bibitemNoStop [0]{.\EOS\space}%
\providecommand \EOS [0]{\spacefactor3000\relax}%
\providecommand \BibitemShut  [1]{\csname bibitem#1\endcsname}%
\let\auto@bib@innerbib\@empty
\bibitem [{\citenamefont {Hagemans}\ and\ \citenamefont
  {Caux}(2007)}]{Hagemans_2007}%
  \BibitemOpen
  \bibfield  {author} {\bibinfo {author} {\bibfnamefont {R.}~\bibnamefont
  {Hagemans}}\ and\ \bibinfo {author} {\bibfnamefont {J.-S.}\ \bibnamefont
  {Caux}},\ }\href {https://doi.org/10.1088/1751-8113/40/49/001} {\bibfield
  {journal} {\bibinfo  {journal} {J. Phys. A: Math. Theor.}\ }\textbf {\bibinfo
  {volume} {40}},\ \bibinfo {pages} {14605} (\bibinfo {year}
  {2007})}\BibitemShut {NoStop}%
\bibitem [{\citenamefont {Vladimirov}(1986)}]{vladimirov1986proof}%
  \BibitemOpen
  \bibfield  {author} {\bibinfo {author} {\bibfnamefont {A.~A.}\ \bibnamefont
  {Vladimirov}},\ }\href {https://doi.org/10.1103/PhysRevA.79.023614}
  {\bibfield  {journal} {\bibinfo  {journal} {Ther. Math. Phys.}\ }\textbf
  {\bibinfo {volume} {66}},\ \bibinfo {pages} {102} (\bibinfo {year}
  {1986})}\BibitemShut {NoStop}%
\bibitem [{\citenamefont {Karbach}\ and\ \citenamefont
  {Müller}(1997)}]{karbach}%
  \BibitemOpen
  \bibfield  {author} {\bibinfo {author} {\bibfnamefont {M.}~\bibnamefont
  {Karbach}}\ and\ \bibinfo {author} {\bibfnamefont {G.}~\bibnamefont
  {Müller}},\ }\href {https://doi.org/10.1063/1.4822511} {\bibfield  {journal}
  {\bibinfo  {journal} {Comput. phys.}\ }\textbf {\bibinfo {volume} {11}},\
  \bibinfo {pages} {36} (\bibinfo {year} {1997})}\BibitemShut {NoStop}%
\end{thebibliography}%

\end{document}


\title{Supplemental Material for \\Asymmetry-induced delocalization transition in the integrable non-Hermitian spin chain}

\author{Yuki Ishiguro$^1$}
\author{Jun Sato$^2$}
\author{Katsuhiro Nishinari$^3$}
 \affiliation{$^1$Department of Aeronautics and Astronautics, Faculty of Engineering, The University of Tokyo, 7-3-1 Hongo, Bunkyo-ku, Tokyo 113-8656, Japan\\
$^2$Faculty of Engineering, Tokyo Polytechnic University, 5-45-1 Iiyama-minami, Atsugi, Kanagawa 243-0297, Japan\\
 $^3$Research Center for Advanced Science and Technology, The University of Tokyo, 4-6-1 Komaba, Meguro-ku, Tokyo 153-8904, Japan
 }

\maketitle


\section{Delocalization of Bethe quantum numbers}
In this section, we present the delocalization of Bethe quantum numbers, which is an exotic property of the string solutions of the asymmetric simple exclusion process (ASEP) induced by non-Hermiticity. 
First, we show the Bethe equations of the ASEP in logarithmic form and introduce the Bethe quantum numbers.
Next, we present the detailed data of the numerical solutions, which are discussed in the main text, and show that the delocalization of the Bethe quantum numbers of the string solutions is allowed in the presence of the non-Hermiticity.

\subsection{Bethe equations in logarithmic form}
Here, we present the Bethe equations of the ASEP in logarithmic form and introduce the Bethe quantum numbers.
The Bethe equations of the ASEP are given by
\begin{align}
            \left[ \frac{1}{\al}\frac{\sin\ze(\la_j+\frac{\im}{2})}{\sin\ze (\la_j-\frac{\im}{2})} \right]^L =  \prod_{\ell \neq j}^N \frac{\sin\ze(\la_j - \la_\ell + \im)}{\sin\ze(\la_j - \la_\ell - \im)} 
            \qquad \text{for} \quad j=1,2,\cdots,N,    
        \label{eq:BAEASEP}
\end{align}
where $L$ is the system size, $N$ is the number of particles, $\alpha := \sqrt{q/p}$, and $\ze := -\log\al$. $p$ and $q$ are hopping rates that satisfy $p\ge q >0$ and $p+q=1$. Since the Bethe equations (\ref{eq:BAEASEP}) have many solutions, it is useful to express them in logarithmic form.
We introduce functions $\theta_n(\la)$  ($n\in \mathbb{N}$) as
\begin{align}
\theta_n(\la) := 
\begin{dcases} 2 \tan^{-1} \left[\coth \left(\frac{n\zeta}{2} \right) \tan \zeta \la \right] + 2 \pi \left\lfloor \frac{\text{Re} (\zeta \la)}{\pi} + \frac{1}{2} \right\rfloor & \text{for} \quad \alpha \neq 1 \\
2 \tan^{-1}\left[\frac{2\la}{n}\right] & \text{for} \quad \alpha = 1,
\end{dcases}
\end{align}
where $\lfloor x \rfloor$ is the floor function defined as $ \lfloor x \rfloor := \max \{n \in \mathbb{Z}|n\le x \}$.
By taking the logarithm of the Bethe equations (\ref{eq:BAEASEP}), we obtain 
\begin{align}
    \theta_1(\la_j) = \frac{2 \pi}{L} I_j + \frac{1}{L} \sum_{\ell \neq j}^N \theta_2 (\la_j-\la_\ell) +\im \log \alpha \qquad \text{for} \quad j=1,2,\cdots,N,
    \label{eq:LogBAE}
\end{align}
where $\{I_j\}$ are the Bethe quantum numbers, which are integers when $L-N$ is odd and half-integers when $L-N$ is even.
Through a set of Bethe quantum numbers, we can characterize the eigenstates of the Hamiltonian.
The total momentum $K$ and the energy eigenvalue $E$ are expressed in terms of Bethe roots $\{\la_j\}$ as
\begin{align}
    &K= \sum_{j=1}^N  [ \pi - \theta_1(\la_j) +  \im \log \al]  = N \pi - \frac{2 \pi}{L} \sum_{j=1}^N I_j,   
    \label{eq:momemtum} \\
     &E= - \frac{1}{2} \tanh \ze \sum_{j=1}^N a_1(\la_j) = 
    \begin{dcases}
        -\sum_{j=1}^N \frac{\Delta \tanh^2 \ze}{\Delta-\cos 2\ze \la_j } & \text{for} \quad \alpha \neq 1 \\
        -\sum_{j=1}^N \frac{2}{4 \la_j^2 +1} & \text{for} \quad \alpha = 1, 
    \end{dcases}
     \label{eq:energy}
\end{align}
where we introduce $\Delta := \cosh \ze$ and functions $a_n(\la)$  ($n\in \mathbb{N}$)
\begin{align}
        a_n(\la):=\frac{d}{d \la} \theta_n(\la) = 
        \begin{dcases}
        \frac{2 \sinh n \ze}{\cosh n\ze -\cos 2\ze \la}    & \text{for} \quad \alpha \neq 1 \\
        \frac{4n}{4 \la^2 +n^2}  & \text{for} \quad \alpha = 1.
        \end{dcases}
\end{align}

\subsection{Numerical results of the two-string solutions}
As mentioned in the main text, we numerically solved the Bethe equations (\ref{eq:BAEASEP}).
Table \ref{tab:NsolData} shows the detailed data for the two-string solutions of the ASEP, which were computed with $1000$ digits precision.
Figure 1 (b) and Fig. 3 in the main text correspond to these data.
From Table \ref{tab:NsolData}, we find the intriguing phenomena on Bethe quantum numbers.

In the Heisenberg spin chain, the Bethe quantum numbers of the string solutions tend to be localized \cite{Hagemans_2007}.
The Bethe quantum numbers of a two-string solution in two-body systems $\{I_j\}=\{I_1, I_2\}$ $(I_1 \le I_2)$ generally satisfy $|I_1 -I_2| \le 1$.
The string solutions with $|I_1-I_2|=1$ are called wide strings, whereas those with $|I_1-I_2|=0$ are called narrow strings.

However, the numerical solutions in Table \ref{tab:NsolData} show that the Bethe quantum numbers of the ASEP might become $|I_1 -I_2| > 1$. 
In the weak asymmetry regime, the difference between the Bethe quantum numbers is $|I_1-I_2|=1$.
As the asymmetry increases, the difference becomes large. 
For example, when the hopping rate is $p=0.6$, the difference is $|I_1-I_2|=3$. 
Thus, the delocalized Bethe quantum numbers are allowed in the presence of the non-Hermiticity.

\begin{table}[hbt]
    \centering
    \caption{Two-string solutions of the ASEP for $L=64$ and $N=2$}
    \begin{tabular}{cccc} 
    \hline
         Hopping rate& Bethe roots  & Energy eigenvalue & Bethe quantum numbers \\ 
         $p$& $\lambda$ &$E$  & $\{I_1,I_2\}$\\\hline \hline 
        0.5 & $1.87069731525+0.500319967464\im$ & $-0.222194559412$& $\{\frac{53}{2},\frac{55}{2}\}$ \\
         & $1.87069731525-0.500319967464\im$ & & \\ \hline
         0.55 & $1.76812968949+0.0795280421456\im$ & $-0.224459345125-0.0831603910095\im$& $\{\frac{53}{2},\frac{55}{2}\}$ \\
         & $1.76832342063-0.919784365696\im$ & & \\ \hline
        0.6 & $1.53829697086-0.211511190590\im$ & $-0.231075006693-0.166256197063\im$& $\{\frac{51}{2},\frac{57}{2}\}$ \\ 
         & $1.53781580701-1.21232109213\im$ & & \\ \hline
          0.65 & $1.29242215650-0.376127957552\im$ & $-0.242257295822-0.249544049466 \im$& $\{\frac{51}{2},\frac{57}{2}\}$ \\ 
         & $1.29348654688-1.37496941921\im$ & & \\ \hline
        0.7 & $1.07594641928-0.458376716696\im$ & $-0.257696964573-0.332315028223\im$& $\{\frac{49}{2},\frac{59}{2}\}$ \\ 
         & $1.07361003425-1.46011069148\im$ & & \\ \hline
        0.75 & $0.893248724590-0.502127184661 \im$ & $-0.277788874400-0.416702538326\im$& $\{\frac{49}{2},\frac{59}{2}\}$ \\ 
         & $0.899574034235-1.49745694729\im$ & & \\ \hline
        0.8 & $0.741344336499-0.514773125989\im$ & $-0.302111990569-0.496520192138\im$& $\{\frac{47}{2},\frac{61}{2}\}$ \\ 
         & $0.726735163156-1.52132494979\im$ & & \\ \hline 
        0.85 & $0.614067763985-0.530731832807\im$ & $-0.322574664876-0.585454599078 \im$& $\{\frac{47}{2},\frac{61}{2}\}$ \\ 
         & $0.616101481608-1.47435909774\im$ & & \\ \hline 
        0.9 & $0.502576797710-0.549017214203\im$ & $-0.329804572647-0.685161545310 \im$& $\{\frac{47}{2},\frac{61}{2}\}$ \\ 
         & $0.548395317443-1.32562771882\im$ & & \\ \hline
        0.95 & $0.389334138196-0.552279568789\im$ & $-0.330820590099-0.779273420786\im$& $\{\frac{47}{2},\frac{61}{2}\}$ \\ 
         & $0.432117780940-1.13705859096\im$ & & \\ \hline
         \end{tabular}
    \label{tab:NsolData}
\end{table}

\section{Complexification of string centers}
In this section, we elucidate the violation of the self-conjugacy, which allows the string centers of the ASEP to be complex.
The Bethe equations of the Heisenberg spin chain, which correspond to Eqs. (\ref{eq:BAEASEP}) for the symmetric case ($p=q$), are given by
\begin{align}
        \left( \frac{\la_j+\frac{\im}{2}}{\la_j-\frac{\im}{2}} \right)^L =  \prod_{\ell \neq j}^N \frac{\la_j - \la_\ell + \im}{\la_j - \la_\ell - \im} \qquad \text{for} \quad j=1,2,\cdots,N.
        \label{eq:BAEXXX}
\end{align}
If $\{\la_j\}$ is a solution of the Bethe equations (\ref{eq:BAEXXX}), its complex conjugate $\{\overline{\la_j}\}$ is also a solution.
The self-conjugacy of the Bethe roots indicates $\{\overline{\la_j}\}=\{\la_j\}$.
Since the Bethe equations of the Heisenberg spin chain exhibit this property \cite{vladimirov1986proof}, the distribution of the Bethe roots is symmetric about the real axis, and the string centers become real.

On the other hand, the Bethe equations of the ASEP ($p\neq q$) do not have this property.
The complex conjugate of Eqs. (\ref{eq:BAEASEP}) are given by
\begin{align}
            \left( \al\frac{\sin\ze(\overline{\la_j}+\frac{\im}{2})}{\sin\ze (\overline{\la_j}-\frac{\im}{2})} \right)^L =  \prod_{\ell \neq j}^N  \frac{\sin\ze(\overline{\la_j} - \overline{\la_\ell} + \im)}{\sin\ze(\overline{\la_j} - \overline{\la_\ell} - \im)} \qquad \text{for} \quad j=1,2,\cdots,N.
            \label{eq:complexBAE}
\end{align}
Thus, $\{\overline{\la_j}\}$, which is a complex conjugate of the Bethe roots, is not a solution to the Bethe equations due to the asymmetry $\alpha$.
In this case, $\{\overline{\la_j}\}$ is a solution to the Bethe equations where the asymmetry is reversed $\alpha \to \alpha^{-1}$. 
Because of the violation of the self-conjugacy, the distribution of the Bethe roots is not symmetric about the real axis. 
Therefore, the string centers of the ASEP are complexified.

\section{Derivation of the Bethe-Takahashi equations}
In this section, we show the detailed procedures for deriving the Bethe-Takahashi equations of the ASEP.
The Bethe-Takahashi equations are reduced Bethe equations in terms of string solutions. 
By considering Bethe roots on a string as a quasiparticle, we derive the equations for the string centers.

We express the Bethe equations of $n$-string solutions (\ref{eq:BAEASEP}) by the string solutions
\begin{align}
     \la_A^{n,j} = \la_A^n + \frac{\im}{2}(n+1-2j) + \delta_A^{n,j} \qquad \text{for} \quad j=1,2,\cdots,n,    
\label{eq:stringsolutions}
\end{align}
as
\begin{align}
            \left[ \frac{1}{\al}\frac{\sin\ze(\laA^{n,j}+\frac{\im}{2})}{\sin\ze (\laA^{n,j}-\frac{\im}{2})} \right]^L =  \prod_{\substack{(m,B) \\ \neq (n,A)}} \prod_{k= 1}^m \frac{\sin\ze(\laA^{n,j} - \laB^{m,k} + \im)}{\sin\ze(\laA^{n,j} - \laB^{m.k} - \im)} 
             \prod_{j'\neq j}^n \frac{\sin\ze(\laA^{n,j} - \laA^{n,j'} + \im)}{\sin\ze(\laA^{n,j} - \laA^{n.j'} - \im)} \quad \text{for} \quad j=1,2,\cdots, n,
        \label{eq:BAEstringform}
\end{align}
where $A$ is the index of the string, $n$ is the length of the string, and $\delta_A^{n,j}$ is the deviation.
The first term of the RHS in Eq. (\ref{eq:BAEstringform}) describes the interactions between the strings, and the second one describes those inside the string. 
The number of particles $N$ are expressed by that of $n$-strings $N_n$ as $\sum_{n=1}^N n N_n =N$.

We reduce the Bethe equations on an $n$-string (\ref{eq:BAEstringform}) by multiplying them as
\begin{align}
          \prod_{j=1}^n  \left[ \frac{1}{\al}\frac{\sin\ze(\laA^{n,j}+\frac{\im}{2})}{\sin\ze (\laA^{n,j}-\frac{\im}{2})} \right]^L = \prod_{j=1}^n \prod_{\substack{(m,B) \\ \neq (n,A)}} \prod_{k= 1}^m \frac{\sin\ze(\laA^{n,j} - \laB^{m,k} + \im)}{\sin\ze(\laA^{n,j} - \laB^{m.k} - \im)}  \prod_{j'\neq j}^n \frac{\sin\ze(\laA^{n,j} - \laA^{n,j'} + \im)}{\sin\ze(\laA^{n,j} - \laA^{n.j'} - \im)}.
        \label{eq:MultipledBAE}
\end{align}
Since the second term of the RHS in Eq. (\ref{eq:MultipledBAE}) satisfies
\begin{align}
        \prod_{j=1}^n \prod_{j'\neq j}^n \frac{\sin\ze(\laA^{n,j} - \laA^{n,j'} + \im)}{\sin\ze(\laA^{n,j} - \laA^{n.j'} - \im)} 
        =1,
\end{align}
Eq. (\ref{eq:MultipledBAE}) is given by
\begin{align}
          \prod_{j=1}^n  \left[ \frac{1}{\al}\frac{\sin\ze(\laA^{n,j}+\frac{\im}{2})}{\sin\ze (\laA^{n,j}-\frac{\im}{2})} \right]^L = \prod_{j=1}^n \prod_{\substack{(m,B)\\ \neq (n,A)}}\prod_{k= 1}^m \frac{\sin\ze(\laA^{n,j} - \laB^{m,k} + \im)}{\sin\ze(\laA^{n,j} - \laB^{m.k} - \im)}.        \label{eq:ReducedMultipledBAE}
\end{align}
Here, we assume that the deviations of the string solutions (\ref{eq:stringsolutions}) vanish, that is $\delta_A^{n,j} \to 0$.
By taking the product over $j$ and $k$ in Eq. (\ref{eq:ReducedMultipledBAE}), we obtain the Bethe-Takahashi equations 
\begin{align}
    \left[ \frac{1}{\al^n} \frac{\sin\ze(\laA^{n}+\frac{\im}{2}n)}{\sin\ze (\laA^{n}-\frac{\im}{2}n)} \right]^L = \prod_{\substack{(m,B)\\ \neq (n,A)}} \varPhi_{nm}(\la_A^{n}-\la_B^{m}),
    \label{eq:BTeqs}
\end{align}
where
\begin{align}
    \varPhi_{nm}(\la):=
    \begin{dcases}
        \frac{\sin \ze (\la+\frac{\im}{2}|n-m|)}{\sin \ze (\la-\frac{\im}{2}|n-m|)} \left[ \frac{\sin \ze [\la+\frac{\im}{2}(|n-m|+2)]}{\sin \ze [\la-\frac{\im}{2}(|n-m|+2)]} \right]^2 \cdots\\ 
          \qquad \cdots
         \left[ \frac{\sin \ze [\la+\frac{\im}{2}(n+m-2)]}{\sin \ze [\la-\frac{\im}{2}(n+m-2)]} \right]^2
         \frac{\sin \ze [\la+\frac{\im}{2}(n+m)]}{\sin \ze [\la-\frac{\im}{2}(n+m)]} \qquad \text{for} \quad n \neq m \\
         \left[ \frac{\sin \ze (\la+\im)}{\sin \ze (\la-\im)} \right]^2  \cdots  \left[ \frac{\sin \ze [\la+\im(n-1)]}{\sin \ze [\la-\im(n-1)]} \right]^2 
         \frac{\sin \ze (\la+\im n)}{\sin \ze (\la-\im n)} 
         \qquad \hspace{10pt} \text{for} \quad n = m.
    \end{dcases}
    \label{eq:BTScattering}
\end{align}

The Bethe-Takahashi equations in logarithmic form are given by
\begin{align}
    \theta_n (\laA^n) = \frac{2 \pi}{L} I_A^n + \frac{1}{L} \sum_{\substack{(m,B)\\ \neq (n,A)}}\Theta_{nm} (\la_A^n-\la_B^m) + \im n \log \al,
    \label{eq:LogBethe-Takahashi}
\end{align}
where 
\begin{align}
    \Theta_{nm} (\la) := 
    \begin{dcases}
        \theta_{|n-m|}(\la) + 2\theta_{|n-m|+2}(\la) +\cdots 
        + 2\theta_{n+m-2}(\la)+\theta_{n+m}(\la) & \text{for} \quad n \neq m \\
         2\theta_{2}(\la) + 2\theta_{4}(\la) \cdots + 2\theta_{2n-2}(\la)+\theta_{2n}(\la) & \text{for} \quad n = m,
    \end{dcases}
    \label{eq:LogBetheTakahashi}
\end{align}
and $\{I_A^n\}$ are the Bethe-Takahashi quantum numbers, which are integers when $L-N_n$ is odd and half-integers when $L-N_n$ are even. 
The total momentum $K$ and the energy eigenvalue $E$ are described in terms of string centers $\{\la_A^n\}$ as 
\begin{align}
    &K = \sum_{n,A} \left[ \pi -\theta_n(\la_A^n) + \im n \log \al \right], \\
    &E=- \frac{1}{2} \tanh \ze \sum_{n,A} a_n(\la_A^n) .
\end{align}


\section{Proof of the classification by the Bethe roots}
In this section, we prove Eq. (9) and Eq. (15) in the main text, which provide the classification of the LHS of the Bethe equations and the Bethe-Takahashi equations in the ASEP.
Since Eq. (9) is a special case of Eq. (15) with $n=1$, we consider the proof of Eq. (15), which is given by
\begin{align}
   \left| \frac{1}{\al^n}\frac{\sin\ze(\la+\frac{\im}{2}n)}{\sin\ze (\la-\frac{\im}{2}n)} \right| \begin{dcases}
        >1 & \quad \text{for} \quad b > c_n(a) \\
        =1 & \quad \text{for} \quad b = c_n(a) \\
        <1 & \quad \text{for} \quad b < c_n(a),
    \end{dcases}
    \label{eq:BTlhscases}
\end{align}
where $n \in \mathbb{N}$, $\la = a +\im b$ ($a$,$b\in \mathbb{R}$), and 
\begin{align}
    c_n(x)=\frac{1}{2\ze} \log \frac{\cos 2\ze x}{\cosh n \ze}.
    \label{eq:centerfunction}
\end{align}

When the hopping rates are symmetric ($\zeta=0$), the ASEP is equivalent to the Heisenberg spin chain. 
In this case, $c_n(x)=0$ and Eq. (\ref{eq:centerfunction}) is obviously satisfied.
In the following, we consider $\ze >0$.
We introduce functions $f_n(x)$ ($n\in \mathbb{N}$) as
\begin{align}
\begin{split}
        f_n(b) &:= \left|\sin\ze \left[a+\im \left(b+\frac{n}{2}\right) \right] \right|^2 -\left|\al^n \sin\ze  \left[a+\im \left(b-\frac{n}{2}\right) \right]\right|^2 \\
        &=e^{-\ze n}\sinh n \ze (e^{2\ze b}\cosh \ze n - \cos2 \ze a).
    \label{eq:functionfn}
\end{split}
\end{align}
The sign of $f_n(b)$ corresponds to the classification of the LHS of Eq. (\ref{eq:BTlhscases}). When $f_n(b)>0$ ($<0$), the LHS of Eq. (\ref{eq:BTlhscases}) is larger (smaller) than one.
From Eq. (\ref{eq:functionfn}), $f_n(b)$ is a monotonically increasing function for $\ze > 0$. The unique solution of $f(c_n)=0$ is given by
\begin{align}
    c_n= \frac{1}{2\ze} \log \frac{\cos 2\ze a}{\cosh n \ze}.
    \label{eq:stringcentercritical}
\end{align}
Therefore, $f_n(b)<0$ for $b<c_n$ and $f_n(b)<0$ for $b>c_n$.
These results lead to Eq. (\ref{eq:BTlhscases}).

\section{Spectra of the Heisenberg spin chain}
In this section, we show the spectra of the Heisenberg spin chain to compare with those of the ASEP. 
The Bethe equations of the Heisenberg spin chain (\ref{eq:BAEXXX}) have two types of solutions: real solutions (one-string solutions) and complex solutions ($n$-string solutions ($n>2$)). 
Real solutions describe scattering states and complex solutions describe bound states.

\begin{figure}[tbh]
    \centering
    \includegraphics[height=6cm]{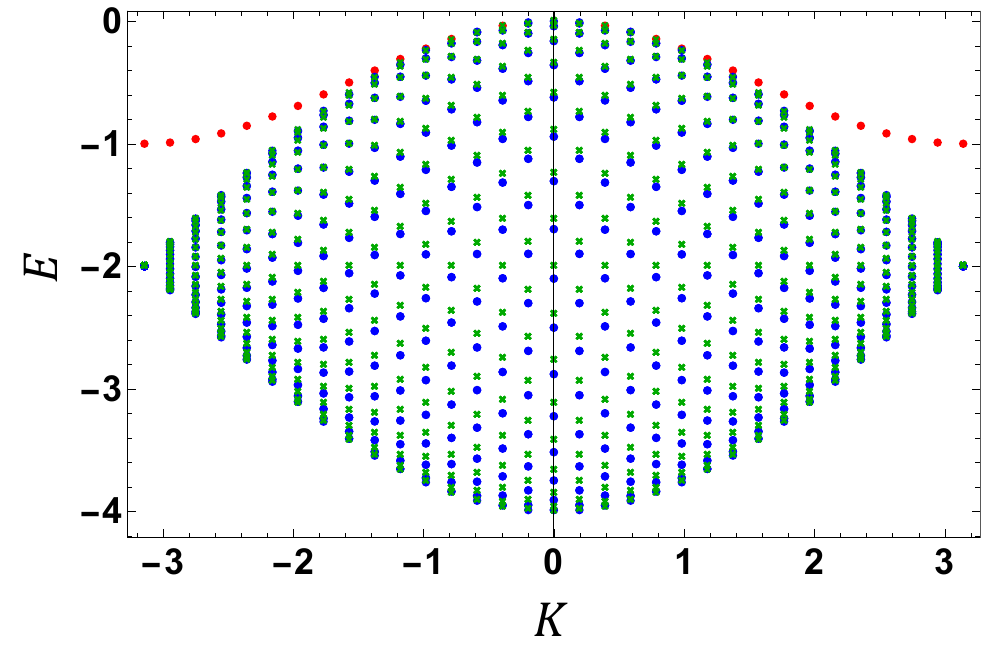}
    \caption{Dispersion relations of the Heisenberg spin chain and the free magnons for $L=32$ and $N=2$. Blue dots show the spectra of real solutions, red dots show the spectra of complex solutions, and green crosses show the spectra of free magnons.}
    \label{fig:dispersion}
\end{figure}

Figure \ref{fig:dispersion} shows the dispersion relation of the Heisenberg spin chain for $L=32$ and $N=2$ \cite{karbach}.
For comparison, we also show the dispersion relation of the free magnons.
The dispersion relation is completely different between real and complex solutions.
The distribution of the spectra of real solutions is similar to that of the free magnons.
The difference between the spectra of real solutions and those of the free magnons is due to the effect of interactions.
Conversely, the distribution of the spectra of complex solutions is different from that of the free magnons. 
Thus, the effect of interactions in the eigenstates of complex solutions is greater than
that of real solutions.

As mentioned in the main text, the ASEP also has two types of spectra: type-I and type-II.
In the strong asymmetry regime, the distribution of type-II spectra blends into that of type-I spectra and becomes hard to distinguish.
This occurred because the effect of the hardcore interactions becomes smaller as the asymmetry increases.

\bibliography{ref_supple}